\documentstyle[11pt,aaspp4]{article}

\tighten
\eqsecnum

\begin{document}
%
\def\kms {km~s$^{-1}$}
\def\hkpc {$h^{-1}$kpc}
\def\hmpc {$h^{-1}$Mpc}
\def\kmsmpc {km~s$^{-1}$ Mpc$^{-1} \,$}
\def\lsim{ \lower .75ex \hbox{$\sim$} \llap{\raise .27ex \hbox{$<$}} }
\def\gsim{ \lower .75ex \hbox{$\sim$} \llap{\raise .27ex \hbox{$>$}} }
\def\items{\hangindent=0.5truecm \hangafter=1 \noindent}
\title{The Cosmological Significance of Disk Galaxy Rotation Curves}

\author{Julio F. Navarro \altaffilmark{1,2}}
\affil{Steward Observatory, The University of Arizona, Tucson, AZ, 85721, USA}
\affil{and}
\affil{Max Planck Institut f\"ur Astrophysik, Karl-Schwarzschild Strasse
1, D-85740, Garching, Germany}
%



\altaffiltext{1}{Bart J. Bok Fellow.}
\altaffiltext{2}{Current address: Department of Physics and
Astronomy, University of Victoria, Victoria, BC, V8P 1A1, Canada.
E-mail: jfn@uvic.ca}


\begin{abstract}
\noindent
We use the rotation curves of more than $100$ disk galaxies to examine
whether the structure of their surrounding dark halos is consistent
with the universal density profile proposed by Navarro, Frenk \& White
(NFW profiles). As noted in previous studies, rotation curve shape is
a strong function of galaxy surface brightness: low-surface brightness
galaxies (LSBs) have slowly rising rotation curves while in
high-surface brightness systems (HSBs) the rotation speed rises
sharply and stays flat or even declines beyond the optical
radius. These observations are in general consistent with NFW halo
profiles, with the possible exception of a few LSBs where the rotation
curves are better described by shallower central density profiles.
Consistency with observational trends requires that halos have lower
characteristic densities than expected in the standard biased cold
dark matter (CDM) scenario, but roughly compatible with
COBE-normalized, low-density, flat CDM universes. The data also imply
that disk mass-to-light ratios increase gently with luminosity,
$(M/L)_{\rm disk} \propto L^{0.2}$, and that the halo circular
velocity, $V_{200}$, is {\it not} directly proportional to the disk
rotation speed, $V_{\rm rot}$.  Slowly rotating disks ($V_{\rm rot}
\, \lsim \, 150$ km s$^{-1}$) are surrounded by halos of higher circular
velocities, whereas faster rotators ($V_{\rm rot} \, \gsim \, 150$
\kms) are all surrounded by halos of similar mass, corresponding to
$V_{200}\sim 200$ \kms. Following a suggestion by Mo, Mao \& White, we
speculate that this is because the efficiency of assembly of baryons
into galaxies is high in massive halos, leading to disks too massive
to be stable in systems that exceed the ``critical'' $V_{200}\sim 200$
\kms. This modeling also provides a natural explanation for the
distribution of sizes and rotation speeds of disk galaxies. Our
results agree well with dynamical studies of binary galaxies and
satellite companions, and help to reconcile discrepancies between the
predictions of semianalytic models of galaxy formation and the
normalization of the galaxy luminosity function and the Tully-Fisher
relation.
\end{abstract}


%
%

\section{Introduction} \label{intro}

Extended rotation curves of disk galaxies provide the most compelling
evidence for the existence of large quantities of dark matter on
galactic scales. Their detailed shapes, in addition, offer invaluable
clues regarding the spatial structure of the dark component and can be
used to gain insight into how the properties of dark halos vary as a
function of the observable properties of the galaxies that inhabit
them.  One well-known example is the apparent coincidence noticed in
early data between the shape of the inner rotation curve and the
circular velocity curve expected from the luminous distribution
alone. This was interpreted as indication that disks dominate the
gravitational potential near the center and that the dark halo plays a
major role only beyond a few exponential disk scalelengths, where the
rotation speed stays roughly constant. Together with the lack of
obvious structural signatures marking the transition between the disk-
and halo-dominated regions, this observation fueled speculation of a
``conspiracy'' between the halo and disk mass distributions that
results in comparable amounts of dark and luminous material within the
galaxy's optical radius (see, e.g., Bahcall \& Casertano 1985, van
Albada \& Sancisi 1986, Sancisi \& van Albada 1987, Freeman 1993).

These early results exerted strong influence on later work, and
subsequent modeling of disk galaxy rotation curves focussed on the
``maximum-disk'' hypothesis, which assigns to the disk the largest
mass-to-light ratio consistent with velocity measurements in the inner
regions. This hypothesis minimizes the halo contribution and imposes
an upper limit on the central density of the halo. One simple halo
model where these constraints are easily implemented is the
non-singular isothermal sphere, which became the model of choice in
mass decomposition studies based on rotation curve data.  This halo
model (referred to hereafter as the ISO model), combined with the
``maximum-disk'' hypothesis, has been found to give remarkably good
fits to the rotation curves of disk galaxies (see, e.g., Begeman 1987,
Broeils 1992, and references therein).

Despite its success in reproducing the observed rotation curves, this
modeling suffers from important shortcomings.  The disk-halo
conspiracy hypothesis, for example, has been challenged by rotation
curves of larger samples of galaxies that include diffuse dwarfs and
compact luminous disks. These larger datasets have made clear that
disk rotation curves come in a variety of shapes, and that only a
subset of spirals have ``flat'' rotation curves. In faint galaxies,
rotation velocities tend to rise slowly and keep rising even beyond
the optical radius, whilst in brighter galaxies rotation speeds rise
fast to a maximum and level off or sometimes even decline beyond the
optical disk. Within the context of the ISO model, a simple (but
non-unique) interpretation of this trend is that halos have large
``core radii'' and that the luminous-to-dark mass ratio inside the
optical radius increases systematically with luminosity. In faint
spirals, disks are relatively unimportant gravitationally and the
rotation curve rises tracing the halo potential. On the other hand, in
bright, massive spirals the contribution of the disk is close to
``maximum'' and results in more steeply rising, roughly flat rotation
curves.

Persic and Salucci (1991) seized on these trends and proposed that the
shape of the rotation curve is determined solely by the luminosity of
the galaxy.  The study of Casertano \& van Gorkom (1991) confirmed the
strong shape-luminosity dependence but noted that galaxies of similar
luminosity have different rotation curve shapes depending on their
surface brightness. At a given luminosity, galaxies with
higher-than-average surface brightness seem to have more steeply
rising rotation curves and to show clearer evidence of the disk
``dominance'' than their lower-than-average surface brightness
counterparts. Together with the fact that luminosity and surface
brightness are correlated (see, eg., Figure 1 below) this suggests
that it is surface brightness rather than luminosity the main
responsible for the shape of the rotation curve.

The hypothesis that disks contribute most of the inner circular
velocity can also be tested by measuring the thicknesses of disks and
their vertical velocity dispersions, which yield an independent
measure of the disk mass. This analysis is perforce statistical in
nature, since it is impossible to measure directly these quantities in
a single galaxy. In all samples where this exercise has been carried
out, it seems that disks contribute only about $60 \%$ of the maximum
observed rotation speed (Bottema 1993, 1997).  

A further difficulty with the ``maximum-disk'' hypothesis is that, if
disks really control the gravitational potential near the center then
two disks of the same mass but different surface density would have
different rotation speeds. This implies that, at fixed luminosity,
disks with higher than average surface brightness should rotate faster
than the mean, and vice-versa (under the plausible assumption that
stellar mass-to-light ratios are independent of galaxy surface
brightness). Such trend would show as a second parameter in the
Tully-Fisher relation, and has been ruled out by detailed
observational studies (Courteau \& Rix 1998).  Unless disk
mass-to-light ratios and surface brightness anti-correlate in exactly
the way needed to cancel the expected trend, these observations
suggest that disks are not ``maximum'' and that halos contribute a
non-negligible fraction of the gravitational potential near the
center.

The theoretical interpretation of ``maximum-disk'' plus isothermal
halo models is also problematic. The halo mass distribution is almost
certainly modified by the collapse of the luminous disk, so it is
difficult to explain how all halos manage to retain their
``isothermal'' shapes {\it after} the disk assembly, especially taking
into account that disks come in many different sizes and masses. If
during the formation of the disk the halo mass distribution reacts
conserving its adiabatic invariants, then the initial halo
configuration must have been hollow, which seems unphysical. The
existence of a core with a well defined central density in the dark
matter distribution is in any case hard to justify, since it implies a
characteristic feature in the power spectrum of initial density
fluctuations on scales where most favored cosmological models would
predict a nearly scale-free behavior.

Finally, non-singular isothermal spheres are poor fits to the
structure of dark halos formed in cosmological N-body simulations, as
illustrated by the recent work of Navarro, Frenk \& White (1996, 1997,
hereafter NFW96 and NFW97, respectively).  Halos formed in
hierarchically clustering universes have density profiles that appear
to diverge near the center and rule out isothermal halos with constant
density cores of the size required to fit the rotation curve data.
The structure of these halos (hereafter NFW profiles, see eq.~4 below)
shows remarkable similarity, regardless of halo mass, power spectrum
of initial density fluctuations, and the value of the cosmological
parameters. As in the isothermal sphere, only two parameters are
needed to describe the halo mass profile: a radial scale (intimately
related to the total mass of the system) and a characteristic
density. One important advantage of the NFW halo model is that the two
free parameters --mass and density-- are now amenable to
interpretation, since they are tightly correlated in each cosmological
model in a way that reflects the collapse times of systems of
different mass. Thus, observational constraints on the characteristic
density of halos of a given mass may in principle discriminate between
competing cosmological models.

The existence of a well-defined prediction for the structure of dark
halos can be used to improve constraints on the relative contribution
of dark and luminous mass in disk galaxies by reexamining the detailed
shapes of their rotation curves.  In this paper, we use NFW halo
profiles to fit the rotation curves of a large sample of galaxies
selected from the literature. The main purpose of this exercise is
twofold: (i) to verify whether NFW profiles are consistent with the
rotation curves of disk galaxies, and (ii) to assess what constraints
these observations place on cosmological models. This analysis should
also enable us to establish quantitatively the relation between galaxy
luminosity and halo mass, a direct indicator of the efficiency of
transformation of baryons into stars in systems of different mass, and
a crucial ingredient of galaxy formation models.

The plan of this paper is as follows. We describe the compilation of
observational data in \S2, and the disk/halo models used to fit the
data in \S3. Section 4 discusses the results of applying these models
to the data, while \S5 discusses our main results in the context of
current cosmological models. A summary of our conclusions is presented
in \S6.

\section{Observational data} \label{obsdat}

We have collected from the literature rotation curves for more than
one hundred disk galaxies. This sample contains most major
compilations of rotation curves available publicly, as well as a
number of individual galaxy studies where the data is presented in
tables that can be cast in electronic form.

\subsection{The sample}

The galaxy sample includes the following. 

(i) Eight galaxies from the PhD thesis of K.Begeman (1987): NGC 3198,
NGC 2403, NGC 6503, NGC 7331, NGC 5371, NGC 2903, NGC 5033, and NGC 2841.

(ii) Eight galaxies from the PhD thesis of A.Broeils (1992): NGC 801,
NGC 1560, NGC 2998, NGC 6674, NGC 2460, NGC 5533, DDO 168, and DDO 105. 

(iii) Sixteen galaxies from the PhD thesis of W.G.de Blok (1997, kindly made
available electronically by the author).

(iv) Nine galaxies from the work of van Zee et al (1997); UGC 191,
UGC 634, UGC 891, UGC 2684, UGC 3174, UGC 5716, UGC 5764, UGC 7178, and
UGC 11820.

(v) Eighty galaxies from the sample of Mathewson et al (1992), as
compiled by Persic \& Salucci (1996, their sample A).

(vi) Twelve galaxies taken from individual studies: DDO 154 from
Carignan \& Beaulieu (1989), DDO 170 from Lake, Schommer \& van Gorkom
(1990), NGC 55 from Puche, Carignan, \& Wainscoat (1991), NGC 247 from
Carignan \& Puche (1990a), NGC 253 from Puche, Carignan, \& van Gorkom
(1991),  NGC 300 from Puche, Carignan, \& Bosma (1990), NGC 2915 from
Meurer, Mackie \& Carignan (1994) and Meurer et al. (1996), NGC 3109
from Jobin \& Carignan (1990), NGC 5585 from Cote, Carignan \& Sancisi
(1991), NGC 7793 from Carignan \& Puche (1990b), UGC 2259 from
Carignan, Sancisi \& van Albada (1988), and IC 2574 from Martimbeau,
Carignan \& Roy (1994).

Data available for all galaxies in this sample include rotation
curves, based on 21cm or H${\alpha}$ observations, as well as
luminosities and scalelengths of the spheroid and disk components.  HI
mass profiles are available for all galaxies with 21cm rotation curve
data, and have been included in the dynamical analyses presented
below.  All luminosities have been scaled to the I-band, using
published data when available in this band, or average colors based on
rough morphological types, as given by de Jong (1995).  Although far
from perfect, these approximate corrections are adequate for the kind
of broad analysis performed in this paper. Distances for all galaxies
are taken from each study or, when unavailable, computed from their
recession velocities and assuming a Virgocentric infall velocity of
300 \kms.  All distances have been scaled to a common value of the
Hubble constant, which we express as $H_0=100 \, h$ \kmsmpc.  The
final sample contains disk galaxies spanning almost four orders of
magnitude in luminosity and three orders of magnitude in surface
brightness, from dwarf irregulars to some of the brightest spirals.

\subsection{Properties of the sample}

The main properties of galaxies in the sample are shown in Figure 1,
where we show correlations between luminosity, exponential disk
scalelengths, average surface brightness, and rotation
speed. Least-squares fits for data in each panel are shown as dotted
lines; the parameters of fits of the form $y=a \log(x)+b$ for all
quantities shown in this figure are listed in Table 1.

The top left panel shows the Tully-Fisher relation, ie. the total
I-band luminosity ($L_I$) versus the maximum rotation speed
($V_{\rm rot}$) measured from the spatially resolved rotation curve. A
single power-law reproduces the data fairly well from the brightest
spirals to the faintest dwarfs in our sample, although there is some
indication of systematic deviations at $V_{\rm rot} \sim 50$ \kms (see
also Meurer et al. 1996).  The bottom left panel shows the correlation
between the luminosity and the exponential disk scalelength
($r_{\rm disk}$). Open circles refer to the total luminosity, and crosses
to the disk luminosity only. The very slight difference between the
two sets of symbols indicates that our sample contains relatively few
galaxies with very massive spheroids.  Because luminosity scales with
radius more steeply than $r^2$ (see Table 1) the average surface
brightness increases with luminosity. This is shown in the top right
panel, where we plot the total luminosity as a function of the
effective surface brightness of the galaxy, $\Sigma_{\rm eff}$, defined
simply as the luminosity of the galaxy divided by the area contained
within one disk exponential scalelength, $\Sigma_{\rm eff}=L_I/\pi
r_{\rm disk}^2$. This overall measure of the central concentration of the
luminous component is straightforward to compute for all galaxies in
the sample and is less sensitive than the central surface brightness
to the presence of a bulge. The correlation between $L$ and
$\Sigma_{\rm eff}$ shown in Figure 1 is similar to that present in galaxy
surveys that are sensitive to low surface brightness galaxies (Impey
\& Bothun 1997) and is therefore unlikely to be solely the result of
selection effects.  Finally, the bottom right panel shows the relation
between the luminosity of the galaxy and the specific angular momentum
of the disk, estimated by $j_{\rm disk} = 2 r_{\rm disk} V_{\rm
rot}$. 

\subsection{Rotation curves}

The rotation curve data compiled from the literature is inhomogeneous
in their spatial coverage and presentation.  The data from Mathewson
et al (1992) consist of H${\alpha}$ velocities along a slit,
de-projected, centered and folded following the procedure outlined by
Persic \& Salucci (1996). The curves are then smoothed by binning the
data in groups of three adjacent data points. The mean velocity is
taken as a measure of the circular velocity at the mean radius of the
bin, and the rms velocity dispersion, weighted by the signal-to-noise
of each measurement, as an indicator of the uncertainty.

Most other rotation curves are derived from 21cm data and are given in
tables of radius vs inclination-corrected rotation speed. Uncertainty
estimates in the rotation speed are also usually quoted, although
there is no consistent practice for determining the size of these
error bars. In some cases formal errors in the average velocity at
each radius from a tilted-ring model are quoted, as given, for
example, by the program ROTCUR (see, eg., Begeman 1987). In other
cases, uncertainties are derived by comparing the approaching and
receding sides of the curve, under the assumption that the curve
should be symmetric relative to the adopted center.

All these estimates are presumably acceptable as rough indicators of
the accuracy of the determination of the velocity at each radius, but
they are subject to different systematics and may result in different
error estimates when applied to the same set of data. For example,
since the error in the mean given by ROTCUR depends on the number of
independent data points available, error bars in the outer parts of
the galaxies, which are well resolved in the velocity maps, are
usually very small ($\sim 1$ \kms). For comparison, the uncertainties
in the velocities estimated by assessing deviations from symmetry of
the folded curves are typically of order $\sim 10$ \kms.

What one would really like for fitting mass models to these data are
estimates of the uncertainty in the {\it circular} velocity at each
radius, something which is in practice much more difficult to
determine. For lack of a better alternative, we have decided to retain
the quoted error bars as estimates of the error in the velocity at
each radius. In practice, this means that minimum-$\chi^2$ values are
not robust measures of goodness-of-fit, and that they should be used
with caution when assessing the quality of fits to two different
galaxies, or when assigning statistical significance to the
quantitative results of the fitting procedure (ie. fits to two
different galaxies may be equally `good' even if the $\chi^2$-values
differ greatly). We shall use the computed $\chi^2$ only to compare
the quality of fits to individual galaxies using different halo
models. This we describe in the following section.

\section{Modeling of rotation curves}

We describe here the two mass models that will be used to fit the
rotation curve data described in the previous section. One is the
usual analytic approximation to the non-singular ``isothermal''
sphere, and the other the NFW profile proposed by NFW96 and NFW97. The
contribution of the luminous component will be taken into account
assuming that the spheroidal component is well described by a de
Vaucouleur's profile, and that the disk is exponential. The parameters
of these components are taken from the studies cited above or, when
unavailable, computed directly from spheroid/disk fits to the surface
photometry. In the interest of simplicity, we shall assume that both
disks and spheroids can be characterized by the same mass-to-light
ratio.  Because our sample contains only a few galaxies where the
spheroid contributes a significant fraction of the luminous budget,
this assumption is unlikely to introduce a severe bias in our
analysis.

\subsection{Halo models}

\subsubsection{The non-singular isothermal sphere}

The approximate fit to the non-singular ``isothermal'' sphere used in
most rotation curve studies is a density profile of the form,
$$
\rho_{{\rm iso}}(r)= {\rho_0 \over 1+(r/r_{c})^2}.  \eqno(1)
$$ 
Its contribution to the circular velocity at each radius is given by,
$$
V_{\rm iso}(r)=4\pi G \rho_0 r_c^2 [1-(r/r_c)\arctan(r/r_c)]. \eqno(2)
$$
The circular velocity increases monotonically with radius and tends to
a constant (maximum) asymptotic velocity,
$$ 
V_{\rm iso}^{\rm max}=\sqrt{4\pi G \rho_0 r_c^2}. \eqno(3)
$$
We shall use $V_{\rm iso}^{\rm max}$ and $r_c$ as the two free
parameters of this model, hereafter referred to as the ISO model.

\subsubsection{The NFW profile}
Spherically averaged density profiles of halos formed in cosmological
N-body simulations of hierarchically clustering universes are well
described by scaling the simple formula proposed by Navarro, Frenk \&
White (NFW96, NFW97),
$$
{\rho(r) \over \rho_{crit}}= {\delta_c \over (r/r_s)(1+r/r_s)^2}, 
\eqno(4)
$$
where $r_s$ is a scale radius, $\delta_c$ is a characteristic
(dimensionless) density, and $\rho_{crit}=3H^2/8 \pi G$ is the
critical density for closure. The associated circular velocity profile
is given by
$$
\biggl({V_c(r)\over V_{200}}\biggr)^2={1 \over x}
{\ln(1+cx)-(cx) /(1+cx) 
\over  \ln(1+c)-c/(1+c)}, \eqno(5)
$$
where $V_{200}$ is the circular velocity at the virial radius,
$r_{200}$, defined as the radius within which the inner mean density
of the halo is $200 \, \rho_{crit}$, $x=r/r_{200}$ is the radius in units
of the virial radius, and $c=r_{200}/r_s$ is the ``concentration''
parameter, a dimensionless number related to the characteristic
density $\delta_c$ by
$$ 
\delta_c={200 \over 3} {c^{3} \over
\bigl[\ln(1+c)-c/(1+c)\bigr]}. \eqno(6)
$$
From these definitions, it follows that
$$
{r_{200} \over h^{-1} {\rm kpc}}={V_{200} \over {\rm km \, s}^{-1}} \eqno(7)
$$
ad that the mass contained within the virial radius is $M_{200}=2.33
\times 10^{5} \, (V_{200}/$\kms$)^{3} \, h^{-1} M_{\odot}$. Thus
halo masses scale with circular velocity not unlike disk luminosities
scale with rotation speed (upper-left panel in Figure 1). This
suggests a natural origin for the Tully-Fisher relation, provided that
disk rotation speeds trace halo circular velocities and that the
stellar mass of the galaxy is approximately proportional to the total
mass of the halo. We shall return to this question in \S5 below.

The density profile of eq.~4 differs significantly from the
``isothermal'' sphere parameterized by eqs.(1) and (2), especially in
its behavior near the center.  $V_c$ increases linearly with radius
for the isothermal sphere but only as $r^{1/2}$ for the NFW
model. Also, whilst the velocity in the isothermal sphere converges
asymptotically to $V_{\rm iso}^{\rm max}$ at large radii, the NFW
circular velocity profile has a maximum at $r\approx 2 \, r_s=2 \,
r_{200}/c$, and declines beyond that radius.  Both models have the
same number of free parameters, a velocity and a radial scale: $V_{\rm
iso}^{\rm max}$ and $r_c$ for the isothermal sphere and $V_{200}$ and
$r_s=r_{200}/c$ for NFW.  We shall use $V_{200}$ and $c$ as the two
free parameters of the NFW model. These two parameters are correlated
for each cosmological model, reducing to one the number of free
parameters of the NFW profile once a cosmological model is adopted. A
simple step-by-step description of how to calculate $c$ as a function
of $V_{200}$ in any hierarchically clustering model is given in the
Appendix of NFW97.
\subsection{The luminous components}
\subsubsection{The exponential disk}
Our modeling assumes that the surface density of the disk is related
to the surface brightness by a single parameter, the disk
mass-to-light ratio $(M/L)_{\rm disk}$, and that it can be approximated by
an exponential form,
$$
\Sigma_{\rm disk}(R)=\Sigma_{\rm disk}^0 e^{-R/r_{\rm disk}}. \eqno(8)
$$ 
Its contribution to the circular velocity on the plane of the disk is then
$$ V_{\rm disk}(R)=V_{\rm disk}^0 x
 {(I_0(x)K_0(x)-I_1(x)K_1(x))^{1/2}}, \eqno(9)
$$
where $V_{\rm disk}^0=(2 \pi G \Sigma_{\rm disk}^0 r_{\rm disk})^{1/2}$ is a
velocity scale, and $I_n$ and $K_n$ are modified Bessel functions of
order $n$.  The circular velocity contribution peaks at
$x=R/r_{\rm disk}\approx 2.2$, where
$$ 
V_{\rm disk}^{\rm max}\approx V_{\rm disk}(2.2 \, r_{\rm disk}) \approx 0.88
\sqrt{\pi G \Sigma_{\rm disk}^0 r_{\rm disk}}, \eqno(10).
$$
and declines rapidly beyond that radius.
\subsubsection{The spheroid}
The spheroidal (bulge) component is assumed spherical and modeled
using de Vaucouleurs' law,
$$
\Sigma_{bulge}(r)=\Sigma_{bulge}^0 \exp{(-r/r_0)^{1/4}} \eqno(11)
$$
The circular velocity contribution of this component is not analytic,
but can be easily computed numerically. For simplicity, we have
assumed that the mass-to-light ratio of the bulge is the same as that
of the disk. In practice this assumption is of little importance in
our analysis, since our sample is heavily biased towards systems with
small bulge-to-disk ratios.
\section{Rotation curve fits}
Fits to the rotation curve of each galaxy are constructed by
straightforward $\chi^2$-minimization varying the three relevant
parameters: the velocity and radial scales of the halo, and the
mass-to-light ratio of the luminous component. Fits using the
``isothermal'' halo model use eq.~2 to specify the contribution of the
dark halo. Fits that use the NFW halo model, on the other hand, take
into account that the dark matter distribution is likely to be
affected by the presence of the luminous component. This is done by
assuming that the disk is assembled slowly so that the adiabatic
invariants of the halo particle orbits are conserved and the halo
responds ``adiabatically'' to the growth of the disk (Barnes \& White
1984, Blumenthal et al. 1986, Flores et al. 1993).  In this
approximation, the radius, $r$, of each halo mass shell after the
assembly of the disk is related to its initial radius, $r_i$, by
$$ r\, \left[ M_{\rm disk}(r)+M_{halo}(r)
\right]=r_iM_i(r_i). \eqno(12) 
$$ 
Here $M_i(r_i)$ is the mass within radius $r_i$ before disk formation
(found by integrating eq.~4), $M_{\rm disk}(r)$ is the final disk mass
within $r$ (found by integrating eq.~8) and $M_{halo}(r)$ is the final
dark matter distribution we wish to calculate. We assume that halo
mass shells do not cross during compression, so that
$M_{halo}(r)=M_{halo}(r_i)=(1-f_{bar})M_i(r_i)$, where $f_{bar}$ is
the assumed initial baryon mass fraction chosen to agree with
primordial nucleosynthesis calculations (Walker et al. 1991, Copi et
al. 1995).

Figure 2 shows the results of the fitting procedure applied to two
different galaxies, a high surface brightness galaxy, NGC 3198
($\Sigma_{\rm eff}=4.2 \times 10^{-2} L_{\odot}/{\rm kpc}^2$, Begeman
1987), and a low surface brightness galaxy, F563-1 ($\Sigma_{\rm eff}=2.2
\times 10^{-3} L_{\odot}/{\rm kpc}^2$, de Blok 1997). 

This figure illustrates a number of features that are common to all
fits. The most important is perhaps to note that rotation curve data
alone cannot be used to discriminate between different halo
models. Indeed, in most cases the minimum $\chi^2$ values obtained
with either halo model are very similar, although the parameters
assigned to each component vary greatly. For F563-1, we find
$(M/L)_{\rm disk}=13.5 \, h \, (M/L)_{\odot}$ using the ISO model and
$(M/L)_{\rm disk}=2.3$ using the NFW halo. The disk dominates the
central potential if an ISO model is adopted but is essentially
negligible if an NFW model is chosen. Differences are less dramatic in
the case of NGC3198, where in neither case the maximum contribution of
the disk (at $2 \, r_{\rm disk}$) exceeds about $\sim 60\%$ of the
measured velocity.

Figure 2 also illustrates clearly that $\chi^2$ values are of little
use taken individually. Because errors in the velocities are estimated
in different ways, the $\chi^2$ is much smaller for F563-1 than for
NGC 3198, most likely reflecting how conservative each author is in
estimating the size of the error bars rather than a true goodness of
fit.
\subsection{Comparison between ISO and NFW halo fits}
Although not meaningful as goodness-of-fit estimators, $\chi^2$ values
obtained using the two halo models described above can still be
compared for individual galaxies. This is illustrated in Figure 3,
where we show minimum $\chi^2$-values for all galaxies in the sample
using the ISO and NFW halo models. Most galaxies can be fit almost
equally well by either the ISO or the NFW model, with a slight
tendency for the ISO model to give smaller $\chi^2$ values. Figure 4
shows that the systems for which the NFW model does worse are
generally low surface brightness galaxies.  The nature of the
discrepancy is illustrated in Figure 5, where we show the rotation
curves of four of the six systems for which ``good'' ($\chi^2 < 1$)
fits can be found for the ISO model but not for NFW halos
($\chi^2>3$). As noted by Flores \& Primack (1994) and by Moore
(1994), these tend to be low-surface brightness dwarf galaxies, where
the HI rotation velocity near the center rises more steeply (and
levels off more sharply) than expected in an NFW profile.

However, it would be premature to conclude that this discrepancy
provides irrefutable evidence against the hierarchically clustering
model on which the NFW profile is based. First of all, it is worth
noting that the discrepancy, although very significant in terms of
$\chi^2$ values, is actually small, as can be appreciated visually in
Figure 5. Indeed, Kravtsov et al (1998) have argued that these data
can be reconciled with cuspy NFW-like density profiles if the inner
asymptotic slope were slightly shallower than proposed by
NFW96. Shallower profiles are actually expected if supernova-driven
winds have played a major role in the formation of these dwarf systems
(Navarro, Eke \& Frenk 1995). Other alternatives include, for example,
the possibility that HI curves are less accurate tracers of the
potential than the formal error bars suggest. Much of the discrepancy
between NFW fits and the data come from the inner few bins, where the
measured velocities are only a few tens of \kms. In this regime,
corrections due to the velocity dispersion of the gas and the finite
thickness of the disk are fairly uncertain and may mask higher
circular velocities than inferred from rotation speeds alone.

A sobering reminder of the sensitivity of rotation curves to adopted
inclinations, signal-to-noise, and other intricacies of the data
analysis process is provided by the case of NGC 3109, where there is a
second published rotation curve which combines Fabry-Perot H${\alpha}$
data in the inner regions with early 21cm data in the outer parts
(solid circles in upper-left panel of Figure 5, data from Carignan
1985). The shape of the rotation curve derived from this independent
dataset is dramatically different from the HI rotation curve, and
quite consistent with an NFW halo model. (An NFW fit with $\chi^2=0.9$
is shown with a dashed line in this panel.)  Finally, preliminary
analysis of optical rotation curves of a large number of low-surface
brightness galaxies shows little evidence for the large discrepancies
with NFW profiles present in this dataset (Pickering et al 1998, in
preparation). This suggests that independent datasets that combine HI
and optical observations are needed in order to assess the true
significance of the discrepancy between this handful of galaxies and
NFW profiles.
\subsection{Rotation curve shapes and NFW halo parameters}
As mentioned in \S1, the parameters of the NFW halo model, $V_{200}$
and $c$, are correlated in a way that depends sensitively on the
cosmological model, so constraints placed on these parameters by
rotation curve data translate directly into constraints on
cosmological models.  It is therefore unfortunate that rotation curves
generally yield only poor constraints on halo parameters. For example,
for the galaxies illustrated in Figure 2, ``acceptable'' fits (as defined
by the condition $\chi^2 < \chi^2_{min} +1$) can be obtained for
$V_{200}$ in the range $(93,510)$ \kms \, and $0.44 < c < 8.3$ in the
case of NGC 3198 and $V_{200}$ in the range $(63,480)$ \kms \, and
$0.01 < c < 19.1$ in the case of F563-1.

The large parameter range allowed by the data reflects the covariance
between parameters in the fit: disk mass can be traded off with halo
concentration to obtain similar circular velocity curves. Actually,
acceptable fits can be obtained for all galaxies in our sample by
setting the disk mass-to-light ratio to zero and varying only the
parameters of the halo.  Figure 6 illustrates this for the case of NGC
3198.  Here we show {\it halo-only} circular velocity curves (eq.~5)
for different values of $c$. ($V_{200}$ is chosen so as to match the
velocity in the outer region and varies by less than 30\% for all the
curves shown.)  Note that the curve labeled $c=26$ is essentially
equivalent to the best-fit shown in Figure 2. It is clear that adding
the disk mass-to-light ratio as a third variable in the fitting
procedure would lead to a large indeterminacy in the value of the
parameters.

Does this mean that it is impossible to derive meaningful cosmological
constraints from rotation curve data? Not necessarily.  One important
thing to note in Figure 6 is that the value of $c$ retrieved from
fitting NFW {\it halo-only} models to the data (hereafter referred to
as $c_{\rm obs}$) represents a firm {\it upper limit} to the
concentration of the halo.  This upper limit applies to a fairly
narrow range of halo circular velocities, which is effectively set by
the rotation velocity in the outer regions of a galaxy under the
plausible assumption that the halo dominates in the outermost
regions. Referring to Figure 6, halos with $c<26$ could in principle
be made consistent with the NGC 3198 data by suitable addition of a
massive disk component, but $c>26$ halos result in rotation speeds
that are already in excess of the data in the inner regions, even
before allowing for the presence of the disk. These halos cannot be
made consistent with the data for any reasonable choice of disk
potential.

A second important thing to note in Figure 6 is that $c_{\rm obs}$ is
a good indicator of the overall shape of the rotation curve. Low
values of $c_{\rm obs}$ ($\lsim 10$) indicate that the rotation curve
rises slowly, and large values of $c_{\rm obs}$ ($\gsim 20$) describe
a sharply rising rotation curve that is flat or that may even decline
in the outer regions. This is because the maximum of the rotation
curve is reached at $r_{\rm max} \approx 2 r_s=2 r_{200} \, c_{\rm
obs}^{-1}=2 (V_{200}/$\kms$) \, c_{\rm obs}^{-1} h^{-1}$ kpc, so the
larger $c_{\rm obs}$ the sharper the rise of the rotation curve and
the nearer the center it reaches the ``flat'' region.  We discuss
below the cosmological implications of the constraints on halo
concentrations imposed by the shape of the rotation curves.
\section{Discussion}
\subsection{Implications for cosmological models}
Upper limits to the halo concentration derived individually for all
galaxies in our sample are shown in Figure 7 as a function of the halo
circular velocity $V_{200}$. This and subsequent figures exclude the
six LSBs which, as discussed in Figure 5, are not adequately fit by
NFW profiles. The arrows indicate the largest $c_{\rm obs}$ of
``acceptable'' (ie. $\chi^2<\chi^2_{min}+1$) fits to the rotation
curve data neglecting the contribution of the disk. Overlaid are the
concentrations expected for halos formed in three cold dark matter
(CDM) cosmogonies.  SCDM refers to the standard biased
($\sigma_8=0.6$)
\footnote{
$\sigma_8$ is the rms mass fluctuations in spheres of radius 8 \hmpc.}
$\Omega=1$ CDM model. The two dotted lines correspond to low-density,
flat ($\Omega+\Lambda=1$) CDM cosmogonies normalized to match the
fluctuations in the cosmic microwave background observed by COBE (see,
eg., Eke, Cole \& Frenk 1996). The scatter around each of these lines
found in N--body simulations is of order $30 \%$ (Navarro et al, in
preparation). The CDM curves are almost horizontal, which implies a
weak dependence between the characteristic density and halo mass. This
is a typical of CDM-like universes, where structure on galaxy scales
grows very fast, and systems of different mass collapse more or less
at the same time. The density (or concentration, see eq.~6) of a halo
traces the density of the universe at the time of collapse, so all
galaxy halos are expected to have similar densities (concentrations).

To be consistent with observations, halos should lie below all upper
limits derived from the rotation curve data. The data in Figure 7
clearly disfavor the SCDM model, but are roughly consistent with the
low-density CDM models shown by the dashed lines. The low
concentrations demanded by the data can only be obtained in
low-density universes, as shown in Figure 8. This figure shows the
concentrations expected for a $200$ \kms \, halo as a function of
$\Omega_0$ and for various choices of the Hubble constant, world
geometry, and power-spectrum normalization. Concentrations of order
$c\approx 3$-$5$ can be obtained for spectra normalized to match the
COBE fluctuations and $\Omega_0 \, \lsim \, 0.3$. However, the dotted
lines (labeled CLUS-$\sigma_8$) show that, in an open ($\Lambda=0$)
universe, COBE-normalized spectra are inconsistent with the
present-day abundance of galaxy clusters (White, Efstathiou
\& Frenk 1993, Eke, Cole \& Frenk 1996). The situation 
improves if a flat geometry ($\Omega_0+\Lambda=1$) is imposed, and
$\Omega_0=0.2$-$0.3$ COBE-normalized spectra are only slightly
inconsistent with the normalization required to fit the cluster
abundance. We conclude that the low concentrations required to fit the
rotation curves of disk galaxies favor low-density, flat models over
open or Einstein-de Sitter CDM universes.

\subsection{Rotation curve shapes and surface brightness}

The galaxies whose rotation curve shapes are inconsistent with the
SCDM model (ie. those with $c_{\rm obs} \, \lsim \, 10$) are
predominantly low-surface brightness galaxies. This is shown in Figure
9, where we plot $c_{\rm obs}$ as a function of the effective surface
brightness of the galaxy, $\Sigma_{\rm eff}$. The two parameters are
highly correlated: low surface brightness galaxies (LSBs) have slowly
rising rotation curves while high-surface brightness disks (HSBs) have
steeply rising, flat rotation curves.  

A simple interpretation for this correlation is that it signals the
increasing gravitational importance of the disk in systems of higher
surface brightness.  Indeed, if disks were gravitationally {\it
unimportant} in all systems, their rotation curves would just trace
the halo potential. The shape parameter $c_{\rm obs}$ would then be
equivalent to the concentration of the halo and would depend very
weakly on surface brightness or other galaxy properties (note the
nearly horizontal lines in Figure 7). The $c_{\rm obs}$-$\Sigma_{\rm
eff}$ correlation suggests that in systems of low surface brightness
the luminous component is indeed unimportant and the rotation curve
traces the mass distribution of the halo, ie. $c_{\rm obs} \sim c
\approx 3$-$5$.  As the surface brightness of the system increases so
does the gravitational importance of the disk. This modifies the shape
of the rotation curve, leading to a steeper inner rise and sharper
flattening that is best described by larger values of $c_{\rm obs}$.

\subsection{Halo masses and disk mass-to-light ratios}

The above discussion implies that we can use the constraints on halo
concentrations derived in the previous section, together with the
observed correlation between $c_{\rm obs}$ and $\Sigma_{\rm eff}$, to
gain insight into disk mass-to-light ratios and the relationship
between the halo circular velocity and the rotation speed of the
disk. In the analysis that follows we shall assume that halos form
with parameters corresponding to the $\Omega_0=0.2$ CDM model shown in
Figure 7. In practice, this means adopting $c \approx 3$ for all
halos, independent of mass. None of the qualitative trends we discuss
below are sensitive to this choice, although quantitatively the
results would change had we adopted $\Omega_0=0.3$ (ie. $c\approx 5$)
instead. The crucial ingredient of the discussion that follows if that
halos of all masses may be described with roughly the same, low ($c\,
\lsim \, 5$) value of the concentration parameter.

What combination of halo masses and disk mass-to-light ratios is
required to reproduce the observed $c_{\rm obs}$-$\Sigma_{\rm eff}$
relation and the correlations shown in Figure 1?  As an illustration,
let us start by assuming that all galaxies have the same stellar
mass-to-light ratio, $(M/L)_{\rm disk}=1 \, h \,
(M_{\odot}/L_{\odot})$ in the I-band. The only free parameter in the
fit is then the halo circular velocity, $V_{200}$, which we fix by
matching the measured maximum rotation speed of the disk within the
radial range covered by the actual data.  The result of this exercise
is shown as dashed lines in Figure 10. Because of our choice of
mass-to-light ratio, low-surface brightness (faint) disks are
gravitationally unimportant, and the rotation curve is dominated by
the halo. The halo circular velocity, on the other hand, does not
reach its virial value within the radii where data are available, and
$V_{200}$ must exceed the rotation speed of the disk in order to match
the Tully-Fisher relation. (Halos reach the maximum circular velocity
at $r_{\rm max} \approx (2/c) (V_{200}/$km s$^{-1}) h^{-1}$ kpc, or
$\sim 66 h^{-1}$ kpc for a $c=3$, $100$ km s$^{-1}$ halo, a radius
well beyond the typical radial extent of rotation curves in dwarf
systems.)  Halos must then be more massive than indicated by a simple
extrapolation of the disk rotation speed, and disks represent a very
small fraction of the total mass, $M_{\rm disk}/M_{200} \, \lsim \, 5
\times 10^{-3}$. This constant mass-to-light ratio model, however,
predicts a much shallower $c_{\rm obs}$-$\Sigma_{\rm eff}$ relation
than observed, as shown in the top-left panel of Figure 10.

A second illustrative example is provided by the dotted lines in
Figure 10, which assume that in all systems the halo circular velocity
at the virial radius is the same as the maximum rotation speed
measured in the disk, ie. $V_{200}=V_{\rm rot}$. The only free parameter
in this case is the disk mass-to-light ratio, which is again adjusted
to match the Tully-Fisher relation. The disk now dominates the central
potential in most systems and the resulting disk mass-to-light ratios
are of order $3$-$5 \, h (M/L)_{\odot}$, in good agreement with the
results of fits that postulate ``maximal disks''. The rotation curve
shapes predicted under this assumption are again in poor agreement
with the data, as shown by the failure of the $c_{\rm
obs}$-$\Sigma_{\rm eff}$ relation to match the observed one.

From these two somewhat extreme examples it is clear that both
$(M/L)_{\rm disk}$ and the relation between $V_{200}$ and $V_{\rm
rot}$ have to vary systematically with galaxy luminosity (or surface
brightness) in order to match the relation between $c_{\rm obs}$ and
$\Sigma_{\rm eff}$. Solid lines in Figure 10 show the result of
adjusting $V_{200}$ and $(M/L)_{\rm disk}$ so as to match the observed
rotation speeds and rotation curve shapes. Two trends are clearly
noticeable. Disk mass-to-light ratios increase systematically with
luminosity, from $\sim 0.5 h \, (M/L)_{\odot}$ in faint, slow-rotating
disks to $2$-$3 \, h \, (M/L)_{\odot}$ in the brightest, fastest
rotators. This represents a modest dependence of $(M/L_I)_{\rm disk}$
on luminosity which can be approximated by $(M/L_I)_{\rm disk} \approx
(L_I/10^9 L_{\odot})^{0.2} h \, (M/L)_{\odot}$. The mass-to-light
ratios of the most luminous disks thus approach those typically
derived for ``maximum disk'' hypothesis (Begeman 1987, Broeils 1992).

The predicted trend of mass-to-light ratio with luminosity is
consistent with the bluer colors of dwarf irregulars compared with
those of bright, high-surface brightness spirals, although the
amplitude of the variation seems slightly larger than can be accounted
for by stellar population synthesis models. For example, assuming star
formation histories that depend exponentially on time and a Salpeter
initial mass function, the models of Bruzual \& Charlot (1993) predict
($B$-$I$) colors of $\sim 1.0$ and $\sim 2.2$ for models with
mass-to-light ratios spanning the range $0.5$-$3$. This is slightly
larger than the color spread from $\sim 1.3$ to $\sim 1.8$ between
late and early type spirals in the data of de Jong (1995), although
his data does not include dwarf irregulars. Metallicity and dust
obscuration are just two of the many effects that would have to be
carefully taken into account before concluding that the mass-to-light
ratios derived from rotation curve fits are indeed inconsistent with
the stellar populations of disk galaxies. Such detailed analysis is,
however, beyond the scope of the present paper.

More intriguing is the resulting relation between $V_{200}$ and
$V_{\rm rot}$. As seen in the top-right panel of Figure 10 (solid
line), halos of disks with $V_{\rm rot}\, \lsim \, 150$ \kms \, have
circular velocities typically higher than $V_{\rm rot}$, by up to $60
\%$ for $V_{\rm rot} \, \lsim \, 100$ \kms. On the other hand, disks
that rotate faster than $\sim 150$ \kms are all predicted to have
similar halo circular velocities, $V_{200} \sim 200$ \kms. In other
words, our results imply that disk-dominated galaxies brighter than
$\sim 5 \times 10^9 h^{-2} L_{\odot}$ are all surrounded by halos of
approximately the same mass, $\sim 2 \times 10^{12} h^{-1}
M_{\odot}$. This behavior has been noted by Persic \& Salucci (1991),
and offers a natural explanation for the puzzling lack of correlation
between luminosity and mass found in dynamical studies of binary
galaxies and of satellites orbiting bright spirals (White et al. 1983,
Zaritsky et al 1993, 1997, Zaritsky \& White 1994). Furthermore, our
results indicate that the lack of correlation arises because these
samples are restricted to fairly luminous galaxies (most galaxies in
the Zaritsky et al sample have $V_{\rm rot} \, \gsim \, 150$
\kms). Extending these studies to fainter galaxies should uncover
evidence that in dwarf systems halos are more massive than indicated
by their rotation speeds and that a clear correlation actually exists
between luminosity and halo mass in faint spirals.
\subsection{Disk galaxy formation in massive halos}
One interesting corollary of the modeling presented in the previous
subsection is that disk galaxies apparently avoid halos more massive
than a ``critical'' circular velocity, $V_{200} \sim 200$ \kms. This
may be related to the mass fraction attached to the luminous galaxy in
systems of different mass. As shown by the solid line in the
bottom-right panel of Figure 10, the disk contributes a negligible
fraction of the total mass in low mass systems, but this fraction
increases steadily as a function of halo mass, exceeding $\sim 10\%$
in halos with $V_{200} \sim 200$ \kms.

The very steep increase in the mass fraction associated with the
luminous galaxy is actually {\it required} by hierarchical galaxy
formation models in order to reconcile the different shapes of the
halo mass function and the galaxy luminosity function (Kauffmann et
al. 1993, Cole et al 1994). For example, according to Cole et al.,
halos with $V_{200} \, \lsim \, 60$ \kms \, should have assembled less
than $1\%$ of their baryons into galaxies. This fraction increases to
$\sim 90\%$ for $V_{200} \sim 200$ \kms, reflecting the strong
dependence of feedback processes on the depth of the potential well
(Navarro \& White 1993). The dramatic increase with halo mass of the
mass fraction attached to galaxies is thus in good agreement with the
independent predictions of semianalytic models of galaxy formation.

The large mass fraction attached to galaxies at the center of massive
halos suggests a simple interpretation for the lack of disk galaxies
in halos that exceed the ``critical'' circular velocity of $\sim 200$
\kms. Disks that dominate the gravitational potential are subject to
global instabilities that can alter profoundly their morphology,
turning them into spheroidal or spheroid-dominated systems
(Efstathiou, Lake \& Negroponte 1982, Christodoulou, Shlosman \&
Tohline 1995). The gravitational importance of disks depends not only
on disk mass but also on size, which is controlled by its angular
momentum: disks with angular momenta below a certain threshold would
therefore be prone to global instabilities.  Mo, Mao \& White (1998)
have recently argued that disks can survive in NFW halos only if their
specific angular momentum, as measured by the rotation parameter
$\lambda${\footnote{The rotation parameter $\lambda$ is defined as $j
|E|^{1/2}/GM^{3/2}$, where $j$ is the specific angular momentum, $E$
is the binding energy, and $M$ is the total mass.}}, exceeds the
threshold $\lambda_{th} \approx f_{\lambda} M_{\rm disk}/M_{200}$,
where $f_{\lambda}$ is a factor close to unity.

Under the conservative assumption that the disk specific angular
momentum does not exceed that of the halo (see, e.g., Navarro \& White
1994, Navarro \& Steinmetz 1997), the fraction of halos that can host
successful disk galaxies can be estimated as the fraction of halos
with rotation parameters higher than $\lambda_{th}$.  The
$\lambda$-distribution of dark halos has been carefully studied
through N--body simulations (Barnes \& Efstathiou 1987, Cole \& Lacey
1996, see also Bartelmann \& Steinmetz 1996), and there is broad
consensus that $\lambda$ has an almost universal distribution that
depends very weakly on halo mass and cosmological parameters. The
median $\lambda$ is $\approx 0.05$, and roughly ninety-per-cent of
systems have $\lambda \, \lsim \, 0.1$. Thus fewer than one in ten
halos with $M_{\rm disk}/M_{200}>0.1$ would be viable hosts of disk
galaxies. Disk mass fractions exceeding $10 \%$ are reached as the
halo circular velocity approaches $V_{200} \sim 200$ \kms (Figure 10),
and turns this into a ``critical'' circular velocity above which disk
survival is unlikely.  

A simple consistency check is possible on this interpretation: the
number density of halos exceeding this critical velocity should be
comparable to the number density of luminous, spheroid dominated
systems in the local universe. It is reassuring that this indeed
appears to be the case. According to Marzke et al (1994), the number
density of E/S0 galaxies brighter than $L_{\star}$ is about $2 \times
10^{-3} h^3$ Mpc$^{-3}$, which compares well with the number density
of systems with $V_{200}>200$ \kms, $\sim 2(3) \times 10^{-3} h^3$
Mpc$^{-3}$, for the $\Omega_0=0.2 (0.3)$ models shown in Figure 7.

We note that the conclusions listed above are sensitive to our choice
of cosmological parameters. Halo number densities, for example, scale
roughly as $\Omega_0$ and would be much higher in an Einstein-de
Sitter universe. The agreement between the abundance of massive halos
and E/S0 galaxies would not hold in an Einstein-de Sitter ($\Omega=1$)
universe. Furthermore, $M_{\rm disk}/M_{200}$ cannot exceed the
universal baryon fraction, $\Omega_b/\Omega_0$, which depends on the
Hubble constant and on the density parameter $\Omega_0$. The
primordial abundance of the light elements suggests that $\Omega_b \,
\sim \, 0.0125 \, h^{-2}$ (Walker et al. 1991, Copi et al 1995), so
$M_{\rm disk}/M_{200}$ may not exceed $\sim 0.05$ in an Einstein-de
Sitter universe if $h\, \gsim \, 0.5$. In other words, unless the
baryon abundance is high, as in the low-$\Omega_0$ universes favored
by rotation curve shapes (baryon mass fractions scale as
$\Omega_0^{-1}$), even systems where most baryons can collect in a
central galaxy would be stable to disk formation.

To summarize, rotation curve shapes imply that the fraction of baryons
that assemble into galaxies is a steep function of halo circular
velocity. The simplest interpretation is that feedback prevents most
baryons from condensing into galaxies in low mass halos. The
effectiveness of feedback decreases rapidly with halo mass, and
baryons collapse unimpeded in systems more massive than about $V_{200}
\, \sim \, 200$ \kms. If the universal baryon fraction is high, as is
the case in the low-density universes favored by the rotation curve
data, global instabilities would prevent the formation of long-lived
disk galaxies in massive systems that exceed this critical circular
velocity.


\subsection{Implications for disk sizes and rotation speeds}

The existence of an upper limit for the mass of halos surrounding
disk-dominated galaxies places strong constraints on disk sizes and
rotation speeds. These two parameters are linked by the specific
angular momentum of the disk which, for simplicity, we shall estimate
as that of an exponential disk with a flat rotation curve, $j_{\rm disk}=2
\, r_{\rm disk} \, V_{\rm rot}$. The simplest assumption is that $j_{\rm disk}$
is proportional to the global specific angular momentum of the halo,
$$ 
j_{\rm disk}=f_j \, j_{halo}=G\, f_j \,
\lambda \, M_{200}^{3/2}/|E|^{1/2}= f_j \, (2/f_c)^{1/2} \,  \lambda \,
(V_{200}/{\rm km \, s}^{-1})^2 \, {\rm km \, s}^{-1} h^{-1} \, {\rm kpc}
\eqno(13), $$
where $f_c\approx (2/3)+(c/21.5)^{0.7}$ if the halo is modeled as an
NFW profile (Mo et al 1997). Since $f_j$ is unlikely to exceed unity
(see Navarro \& Steinmetz 1997, and references therein), and very few
halos are expected to have $\lambda > 0.1$, the condition
$V_{200} \, \lsim \, 200$ \kms \, effectively translates into a
maximum specific angular momentum for disks, $j_{\rm disk}^{\rm max}
\sim 6\times 10^3$ \kms $h^{-1}$ kpc. This is in good agreement
with the highest specific angular momentum observed in our sample,
$j_{\rm disk}= 5.3 \times 10^{3}$ \kms $h^{-1}$ kpc, as shown in
Figure 11. \footnote{We note that Malin I, the highest specific
angular momentum disk known, has $j_{\rm disk}
\sim 9.5 \times 10^{3}$ \kms $h^{-1}$ kpc (Pickering et al 1997), and
therefore exceeds our limit by about 60\%. If our interpretation is
correct, Malin I would correspond to a $\lambda \sim 0.2$ halo, and
should be a very rare system indeed. The discovery of a large number
of systems like Malin I would seriously undermine the validity of the
conclusions reported here.}  Figure 11 shows data for all galaxies in
our sample with open circles, and has been augmented by the galaxies
reported by Courteau (1997), shown as filled circles.

In this interpretation, galaxies with $j_{\rm disk} \sim j_{\rm
disk}^{\rm max}$ are approximately ``maximal rotators'', ie. disks
assembled in the highest-$j$ ($\lambda=0.1$), most massive ($V_{200}
\sim 200$ km s$^{-1}$) halos without any loss of angular momentum
($f_{j} \sim 1$).  With this assumption the characteristic size of
these disks can be estimated directly. Following Mo et al (1997, see
their eq.28), disk exponential scalelenghts are directly proportional
to $\lambda$ and the halo virial radius, modulated by factors that
depend on the concentration of the halo, on the mass of the disk and
on the available angular momentum,
$$ r_{\rm disk}=(2f_c)^{-1/2} \, f_j \, f_r \, \lambda\, r_{200},
\eqno(13) $$
where the dimensionless function $f_r(\lambda,c,M_{\rm disk}/M_{200},f_j)$
can be approximated by
$$
f_r\approx(f_j \lambda/0.1)^{-0.06+2.71m_d+0.0047/(f_j \lambda)}
(1-3m_d+5.2m_d^2)(1-0.019c+0.00025c^2+0.52/c), \eqno(14)
$$
and we have defined $m_d=M_{\rm disk}/M_{200}$.  In ``maximal
rotators'' the mass fraction $m_d$ approaches the maximum allowed by
the universal baryon fraction, $m_d \sim 0.0125 h^{-2}/\Omega_0 \,
\lsim \,0.25$ (assuming $h\, \gsim \, 0.5$ and $\Omega_0 \, \gsim \,
0.2$) . Using $m_d=0.2$, $f_j=1$, $\lambda=0.1$, and $V_{200}=200$
\kms \, we derive a rough upper limit  $r_{\rm disk}$ of $\sim 10 \,
h^{-1}$ kpc.

In order to derive the dependence of this maximum radius on disk
rotation speed, we need to know how the halo virial radius (or its
circular velocity $V_{200}$) scales with $V_{\rm rot}$. This is shown
in the upper-right panel of Figure 10 for the data in our sample
(assuming $\Omega_0=0.2$, or $c \sim 3$), and can be parameterized by
\footnote{The corresponding relation for $\Omega_0=3$ ($c\sim 5$) can
be obtained by replacing ($9/4, 300$ km s$^{-1}$) for ($3/2, 500$ km
s$^{-1}$) in eq.(15).},
$$ V_{200}={9 \over 4} {V_{rot} \over (1+V_{rot}/300 \, {\rm km \,
s}^{-1})^{7/4}} \eqno(15)
$$
Using eq.(15), the maximum $r_{disk}$ scales with rotation speed as
shown by the solid line labeled $\lambda=0.1$ in Figure 11. This line,
together with an analogous one computed using the effective minimum in
the $\lambda$-distribution ($\lambda \sim 0.02$) should bracket all
observed disk galaxies if our interpretation is correct.  This is in
good agreement with the data analyzed in this paper and those of
Courteau's sample, as can be seen in Figure 11.

As mentioned in \S5.4, the distribution of data points in the $r_{\rm
disk}$-$V_{\rm rot}$ plane is further constrained by global
instabilities which prevent disks from forming in systems with
rotation parameters below a certain threshold,
$\lambda_{th}=f_{\lambda} m_d$. Mo et al (1997) advocate $f_{\lambda}$
of order unity, although newer calculations by Syer, Mao \& Mo (1997)
and Sellwood \& Moore (1998) argue for significantly smaller values of
$f_{\lambda}$. Within the context of our discussion, $f_{\lambda}$ is
effectively determined by our assumption that the highest angular
momentum disks in our sample are maximal rotators, which implies
$f_{\lambda} \sim 0.5$.

The actual constraint imposed by $\lambda_{th}$ on the $r_{\rm
disk}$-$V_{\rm rot}$ plane depends sensitively on how $m_d$ scales
with halo mass. For example, if all galaxies had the same (maximum)
$m_d \sim 0.2$, then $\lambda_{th}=0.1$ and all disks would lie on the
solid line labeled $\lambda=0.1$. On the other hand, a very steep
dependence of $m_d$ on $V_{200}$, such as that derived in the previous
subsection (see solid line in the bottom-right panel of Figure 10),
implies an upper limit on rotation speeds of $\sim 350$ \kms \, that
is essentially independent of disk radius. This is shown by the curve
labeled $\lambda_{th}$ in Figure 11.

Our interpretation thus predicts that the distribution of
disk-dominated galaxies in the $r_{\rm disk}$-$V_{\rm rot}$ plane
should be confined to within the ``wedge'' determined by the three solid
lines in Figure 11. This implies that the higher the angular momentum
the narrower the scatter in the properties of disks, such as rotation
speed and scalelenght. This ``convergence'' in the properties of
high-$j_{\rm disk}$ disks seems to apply also to their luminosities:
the scatter in luminosity at a fixed $j_{disk}$ decreases towards
higher luminosities (see bottom-right panel in Figure 1). Since
luminosity and angular momentum scale with distance in different ways,
it may be worth investigating whether the spin of bright disk galaxies
could actually be used as an accurate distance indicator.

By contrast, the assumptions $V_{\rm rot}=V_{200}$ and 
$m_d=$constant allow galaxies to form anywhere between the two dotted
lines in this figure. (The two lines correspond to $m_d=0.15$,
$\lambda=0.1$ and $\lambda=0.02$, respectively.)  Galaxies with
rotation speeds above $350$ \kms \, and radii above $\sim 10 h^{-1}$
kpc are permitted in this scenario but are not observed. Our modeling
thus offers a natural explanation for the existence of an abrupt
cutoff in the distribution of disk rotation speeds, a puzzling
observation that has so far eluded proper interpretation (Peebles \&
Silk 1990).


\section{Summary and Conclusions}

We have analyzed the rotation curves of more than 100 disk galaxies
taken from the literature in order to examine whether the structure of
their surrounding dark halos is consistent with the NFW model proposed
by Navarro, Frenk \& White on the basis of cosmological N--body
simulations (NFW96, NFW97). Our main conclusions can be summarized as
follows.

1) The rotation curves of disk galaxies, with the possible exception
of a few low-surface brightness dwarf irregulars, are in general
consistent with NFW mass profiles. The disagreement with the rotation
curves of the dwarfs, although significant, is not large, and may
signal either systematic departures from the NFW shape in dwarf galaxy
halos such as those proposed by Kravtsov et al (1998), or systematic
deviations between measured HI velocities and circular velocities near
the center of these systems.

2) The overall shape of the rotation curves can be used to place
strong constraints on the characteristic concentrations of halos
surrounding disk galaxies.  We find that consistency with observations
implies that halos must be less centrally concentrated than expected
in the standard CDM scenario, but are roughly consistent with
low-density ($\Omega_0\, \lsim \, 0.3$), flat CDM cosmogonies
normalized to match the CMB fluctuations measured by the COBE
satellite.

3) Rotation curve shape correlates strongly with surface brightness:
low-surface brightness galaxies have slowly rising rotation curves,
while their high-surface brightness counterparts have sharply rising,
flat, or even declining rotation curves. 

4) This trend has important implications on disk mass-to-light ratios
and halo masses.  I-band disk mass-to-light ratios are predicted to
increase weakly but systematically as a function of galaxy luminosity,
from $(M/L)_{\rm disk} \sim 0.5 h M_{\odot}/L_{\odot}$ in dwarf
irregulars to $\sim 3 h M_{\odot}/L_{\odot}$ in the brightest
disk-dominated spirals. This trend is consistent, although perhaps a
bit too pronounced, with the systematic blueing of disks in the Hubble
sequence towards fainter, more irregular galaxies.

5) Halo circular velocities are not simply proportional to disk
rotation speeds. Halos of faint galaxies have circular velocities that
can exceed by up to $60\%$ the rotation speed of the disks, and the
opposite seems to hold in the brightest spirals. Disks with rotation
speeds in the range ($150$, $350$) \kms \, appear to be surrounded by
halos of $\sim 200$ \kms. This explains naturally the weak dependence
between halo mass and luminosity found in dynamical studies of binary
galaxies and satellite-primary pairs. It also implies the existence of
a critical halo mass above which disk galaxies would be too massive to
survive as such.

6) The above conclusions pose important constraints on the
distribution of disk sizes and rotation speeds. Together with general
considerations regarding global disk stability, they impose firm upper
limits on the angular momentum, size, and rotation speed of disks. In
particular, they predict that very few galaxies with $r_{\rm disk} \,
\gsim \, 10 h^{-1}$ kpc, $V_{\rm rot} \, \gsim \, 350$ \kms, and
$j_{\rm disk} \, \gsim \, 6 \times 10^3$ \kms $h^{-1}$ kpc should
exist, in good agreement with observations.

Our results have important applications in semianalytic models of
galaxy formation. They confirm independently the strong dependence on
halo circular velocity of the mass fraction attached to luminous
galaxies that had been hypothesized in order to reconcile the the dark
halo mass function with the galaxy luminosity function. Our modeling
also suggests that semianalytic models that assume a direct
proportionality between $V_{200}$ and $V_{\rm rot}$ may have
substantially overestimated the number of $L^{\star}$ galaxies. The
relation between $V_{200}$ and $V_{\rm rot}$ suggested by our modeling
(Figure 10 and eq.15) implies that fewer luminous galaxies are
expected, thus offering a promising explanation for the failure of
semianalytic models to match simultaneously the luminosity density of
the universe and the zero point of the Tully-Fisher relation
(Kauffmann et al. 1993, Cole et al. 1994).

Our analysis also raises a number of questions that should be explored
in further studies. For example, why is the efficiency of
transformation of baryons into stars such a steep function of halo
circular velocity?  Are the highest angular momentum disks really
maximal-rotators? Are low-surface brightness dwarfs really consistent
with NFW-like halo models? And above all, what is the origin of the
Tully-Fisher relation? Our analysis has made use of this surprisingly
tight relation between disk luminosity and rotation speed, but
provides no firm clues to elucidate its origin. Actually, by pointing
to a non-linear relation between halo circular velocity and disk
rotation speed it questions the most promising scenario: that the
Tully-Fisher relation just reflects the equivalence of mass and
circular velocity ($M \propto V^3$) in systems formed within a
cosmological context. Until a clear understanding of this and other
strikingly tight correlations between the dynamical, morphological,
and luminous properties of disk galaxies emerges, our grasp of the
processes that formed and shaped galaxies is bound to remain
incomplete.

%


\acknowledgments
I am grateful to Hans-Walter Rix, Houjun Mo, Carlos Frenk, Simon
White, and Mike Hudson for useful discussions and to Paolo Salucci,
Liese van Zee, Erwin de Blok, and Stephane Courteau for kindly making
available data in electronic form.
%
%
%
%
%
%
%

\clearpage



\begin{table*}
\begin{center}
\begin{tabular}{crrrr}
& $L_I$-$V_{\rm rot}$ 
& $L_I$-$\Sigma_{\rm eff}$ 
& $L_I$-$r_{\rm disk}$
& $L_I$-$j_{\rm disk}$ \\
\tableline
$a$ & 0.26 & 0.39 & 0.30 & 0.55 \\ 
$b$ & 2.29 & -1.58 & 0.54 & 2.90  \\ 
$Y$-rms & 0.08  & 0.36  & 0.18  & 0.19  \\
         &       &       &       &       \\
 $a'$    & 3.27  & 1.02  & 2.03  & 1.49  \\
 $b'$    & -7.56 & 1.35  & -1.27 & -4.40  \\ 
$X$-rms & 0.29  & 0.57  & 0.46  & 0.31	  \\
 \end{tabular} 
\end{center} 
\caption{
Parameters of the least-squares fits shown in Figure 1. For each
$X$-$Y$ combination we show the parameters of fits of the form
$\log(Y)=a \log(X) + b$ and $\log(X)=a' \log(Y) + b'$.} \label{tbl-1}
\end{table*} 

\vfill\eject

\clearpage 
\begin{figure}
\plotone{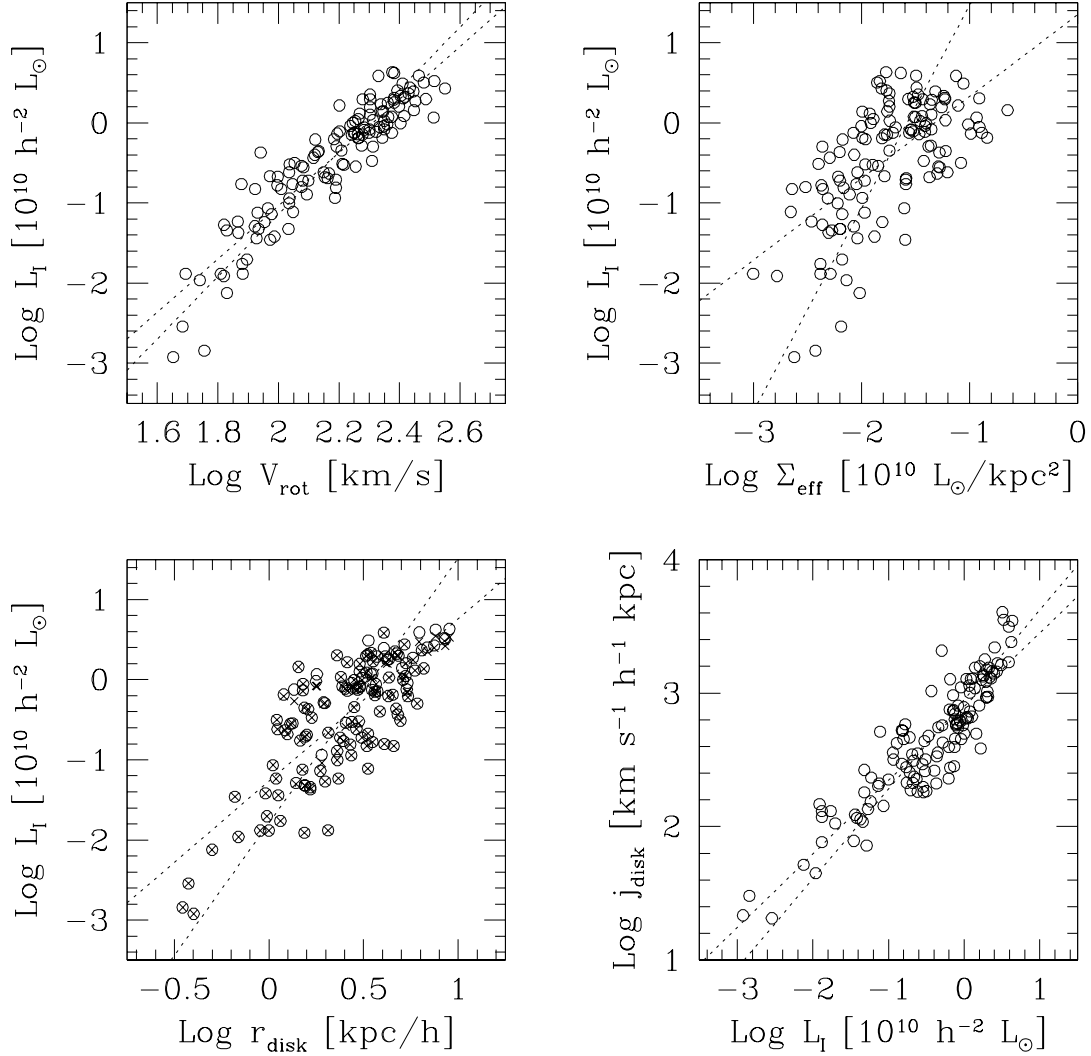}
\figurenum{1} 
\caption{
Correlation between global properties of galaxies in the sample. $L_I$
is the total I-band luminosity, $V_{\rm rot}$ is the maximum measured
rotation speed, $r_{\rm disk}$ is the exponential disk scalelength,
$\Sigma_{\rm eff}=L_i/\pi r_{\rm disk}^2$ is the ``effective'' surface
brightness, and $j_{\rm disk}$ is the disk specific angular
momentum. Units are given in the labels.}
\end{figure}  
\begin{figure} 
\plotone{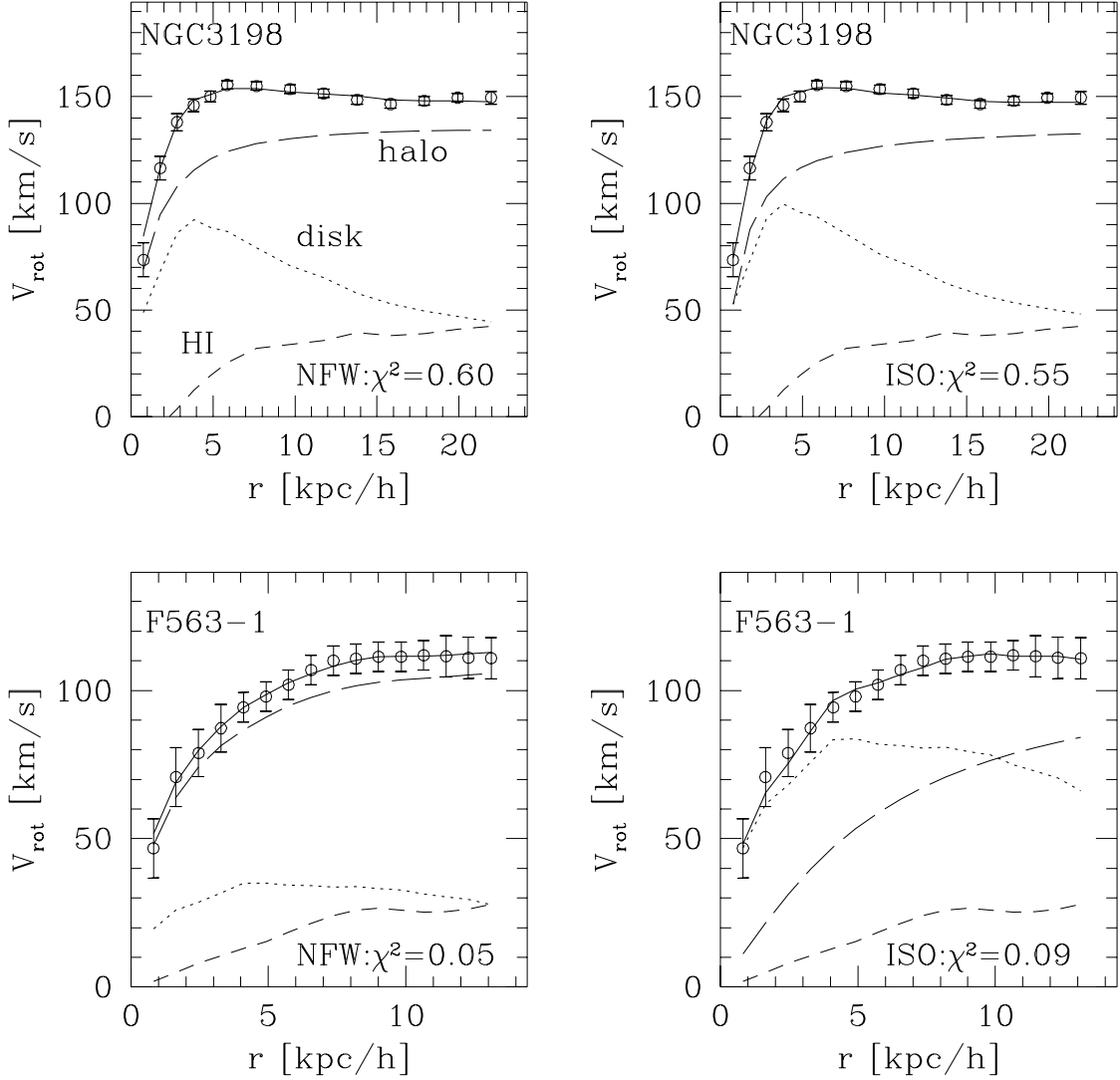}
\figurenum{2} 
\caption{
Rotation curve fits using the NFW and the ISO halo models shown for a
high-surface brightness galaxy (NGC 3198, Begeman 1987) and a
low-surface brightness galaxy (F563-1, de Blok 1997). Note that either
halo model produces acceptable fits, although they may require
different disk contributions.}
\end{figure}  
\begin{figure} 
\plotone{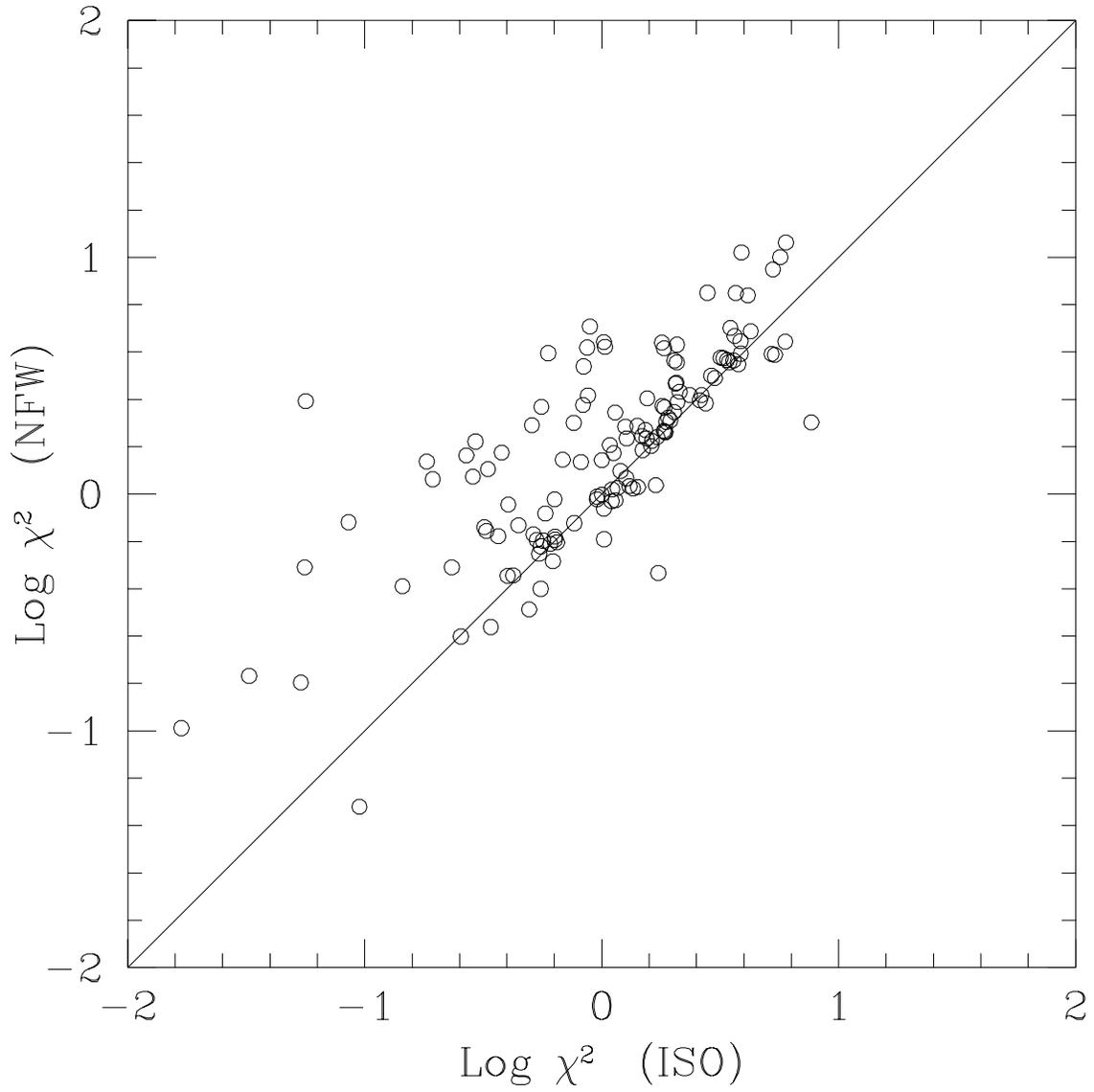}
\figurenum{3} 
\caption{
Minimum reduced $\chi^2$-values obtained by fitting NFW or ISO models
to the rotation curves of galaxies in our sample. See text for
details.  }
\end{figure}  
\begin{figure} 
\plotone{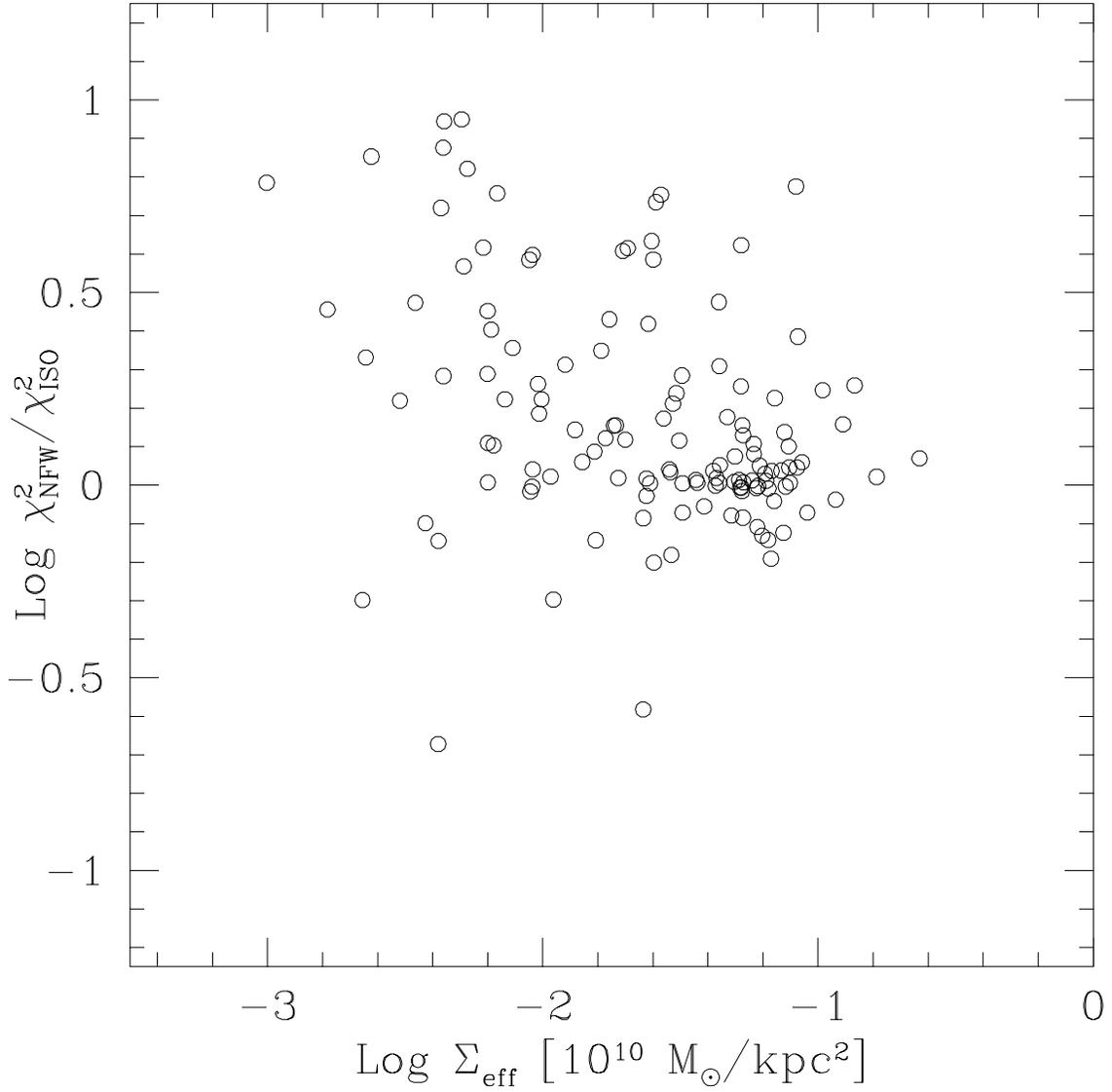}
\figurenum{4} 
\caption{
Ratio of minimum $\chi^2$-values obtained by fitting NFW and ISO
models to the data, plotted as a function of effective surface
brightness. Note that low surface brightness galaxies are in general
better fit with the ISO model. }
\end{figure}  
\begin{figure} 
\plotone{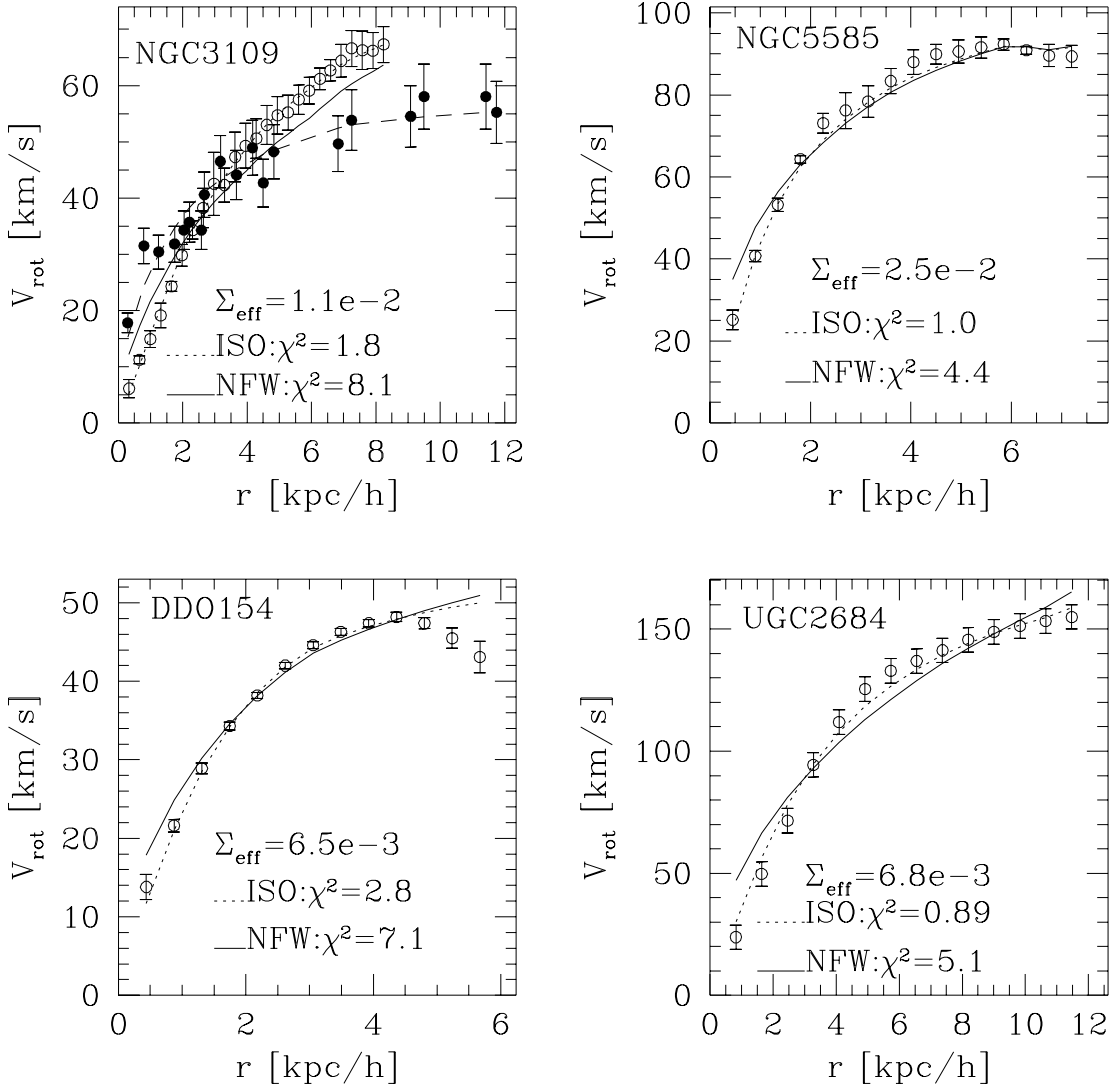}
\figurenum{5} 
\caption{
Same as Figure 2, but for four low-surface brightness galaxies that
are better fit using the ISO halo model rather than an NFW
profile. For NGC 3109 two sets of data are shown. The HI rotation
curve of Jobin \& Carignan (1990, open circles) and a combined
H${\alpha} +$ HI rotation curve of Carignan (1985, filled
circles). These two sets of data data disagree significantly, and
illustrate the danger of over-interpreting the apparent failure of the
NFW halo model, especially since such model can reproduce the
H${\alpha} +$ HI data quite well (dashed line).}
\end{figure}  
\begin{figure} 
\plotone{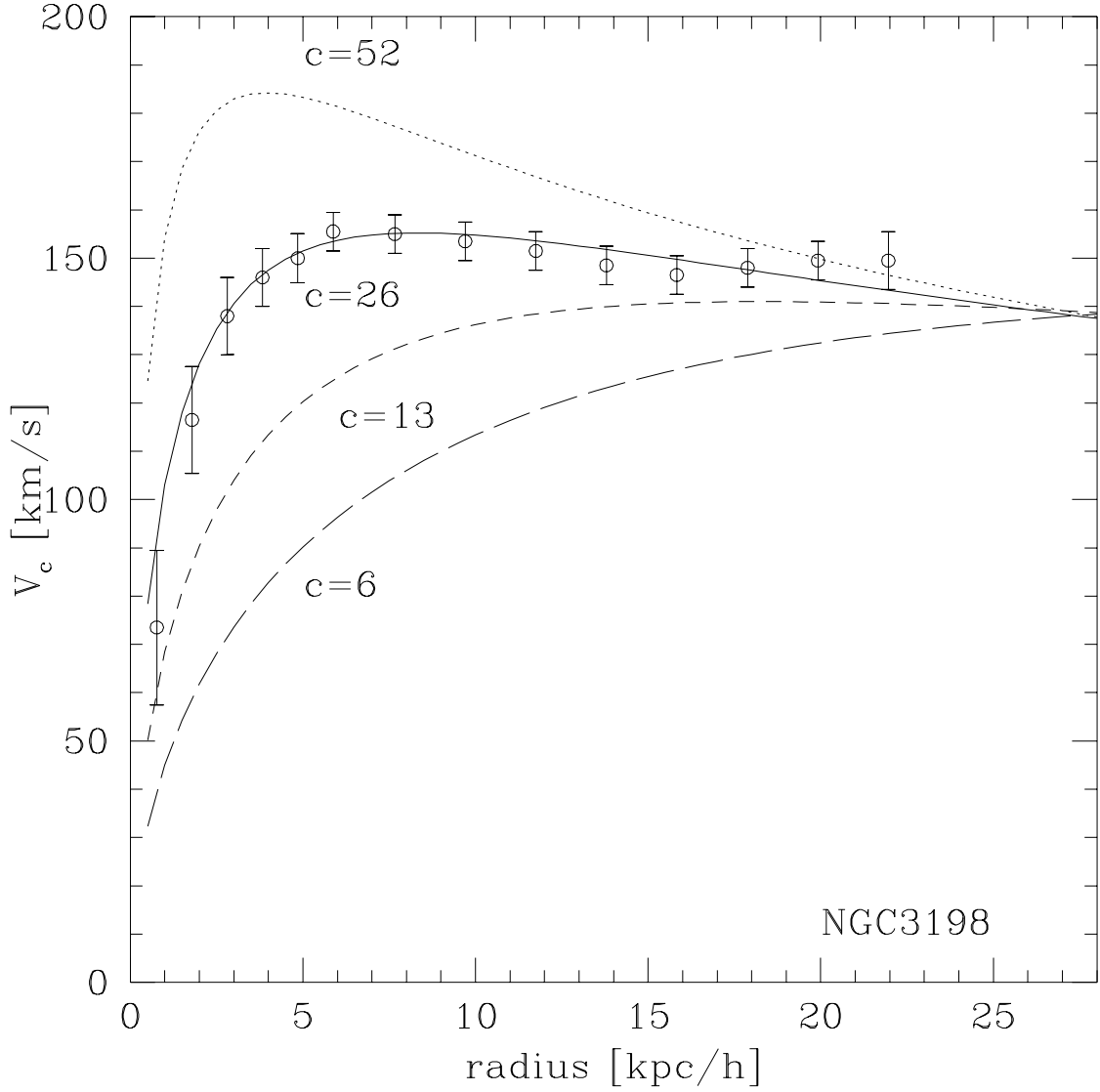}
\figurenum{6} 
\caption{
NFW circular velocity profiles (eq.~5) for different values of the
concentration. The circular velocity at the virial radius, $V_{200}$,
is chosen to match the rotation speed in the outskirts of the galaxy,
and varies from $100$ to $130$ \kms \, from top to bottom. As
illustrated in this figure for NGC 3198, most galaxy rotation curves
can be adequately parameterized by eq.~5 alone. The best fit value
of the $c$ parameter ($c_{\rm obs} \approx 26$ in this case) provides
a quantitative measure of the shape of the rotation curve. Low values
of $c_{\rm obs}$ denote slowly rising curves, while high values of
$c_{\rm obs}$ indicate steeply rising, flat or declining rotation
curves. Furthermore, $c_{\rm obs}$ constitutes a firm upper limit to
the concentration of the halo, since the fit neglects the contribution
of the luminous component.}
\end{figure}  

\begin{figure} 
\plotone{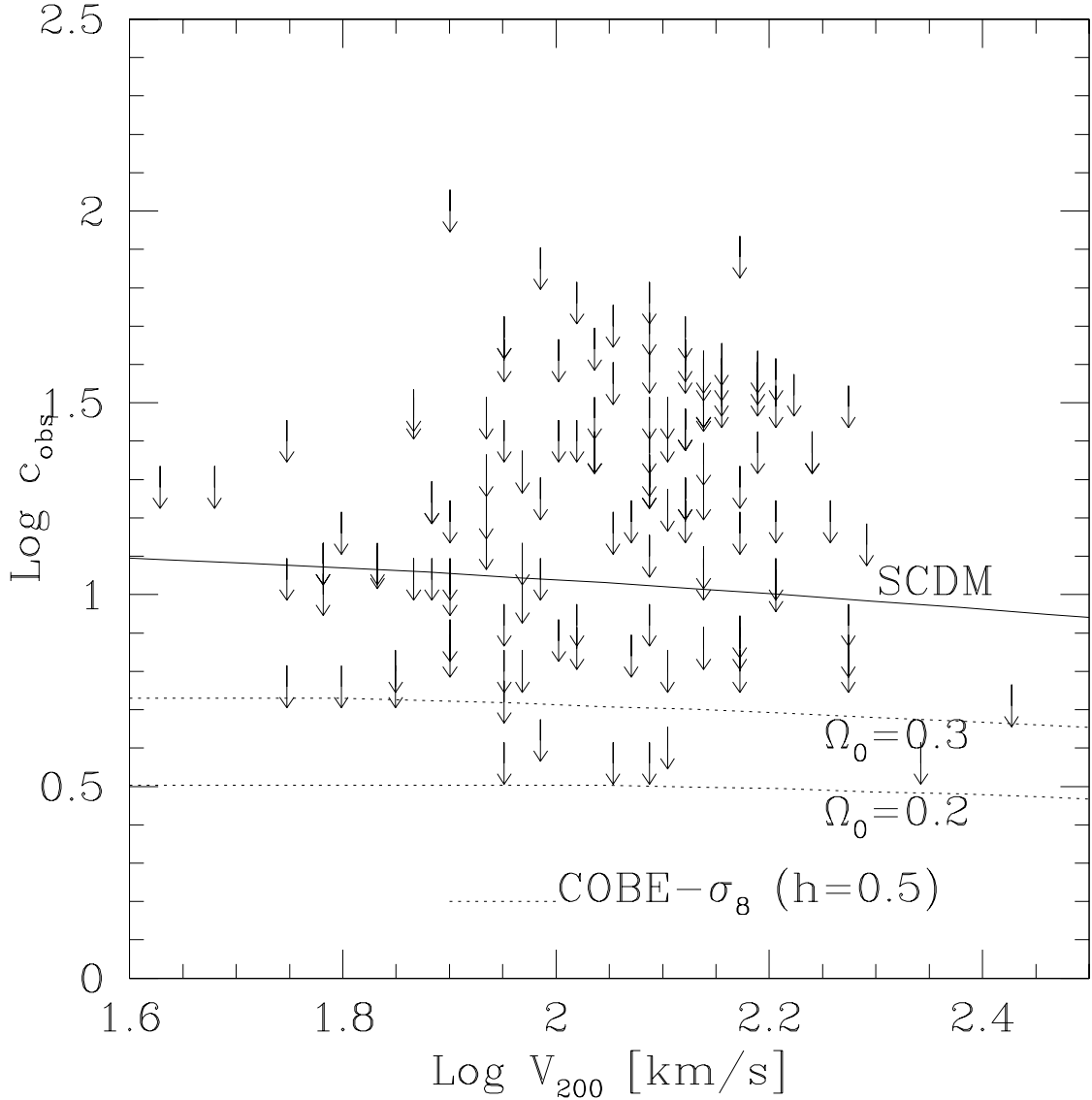}
\figurenum{7} 
\caption{
Upper limits to halo concentrations derived for all galaxies in the
sample, as a function of halo virial velocity. The nearly
horizontal lines are the predictions of NFW97 for three different
cosmological models: the standard biased CDM model ($\Omega=1$,
$\sigma_8=0.6$) and two low-density flat CDM models ($\Omega+
\Lambda=1$) normalized to match the CMB temperature fluctuations
measured by COBE.  The data clearly indicate that low values of the
concentration ($c \, \lsim \, 5$) are required for consistency with
observed rotation curves. This is inconsistent with standard CDM and
favors the $\Omega<1$ models. }
\end{figure}

\begin{figure} 
\plotone{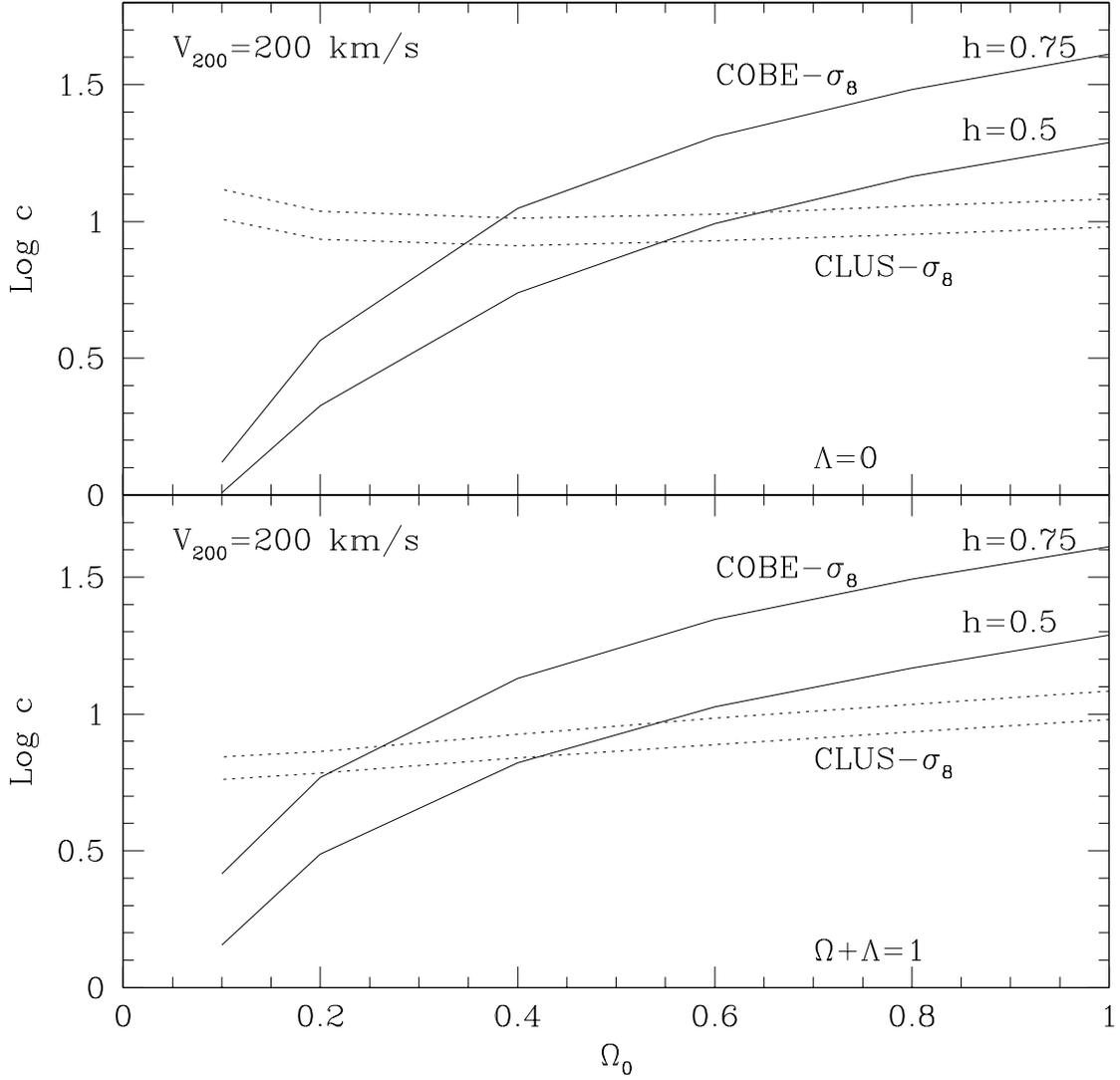}
\figurenum{8} 
\caption{
Predicted concentrations of a $V_{200}=200$ \kms \, NFW halo as a
function of $\Omega_0$ for various choices for the Hubble constant and
the cosmological constant (see NFW97). Solid lines have been
normalized to match the CMB fluctuations measured by COBE; dotted
lines are normalized to match the abundance of rich galaxy clusters
following Eke et al. (1996). Top lines correspond to $h=0.75$, bottom
lines to $h=0.5$. Note that the low-values of $c$ ($\lsim \, 5$)
required to match the data (Figure 7) can only be obtained in
low-density, flat CDM models. The cluster normalization appears to
exclude open models. $\Omega_0=1$ models also appear to be ruled
out. }
\end{figure}

\begin{figure} 
\plotone{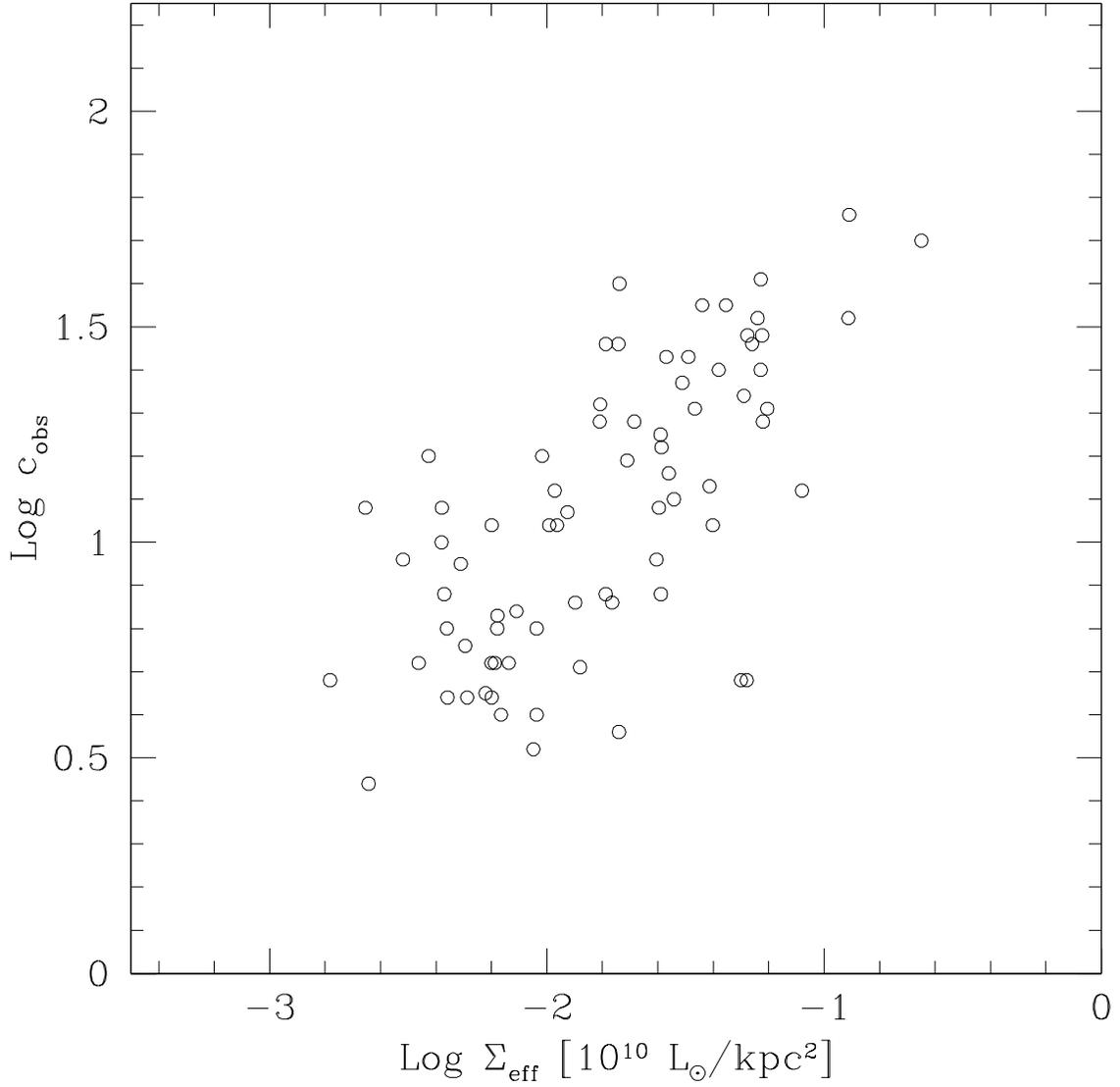}
\figurenum{9} 
\caption{
The rotation curve ``shape'' parameter $c_{\rm obs}$ plotted as a function
of the effective surface brightness $\Sigma_{\rm eff}$. A clear trend
is noticeable: LSBs have slowly rising rotation curves whilst HSBs
have steeply rising, flat or sometimes even declining rotation
curves.}
\end{figure}  

\begin{figure} 
\plotone{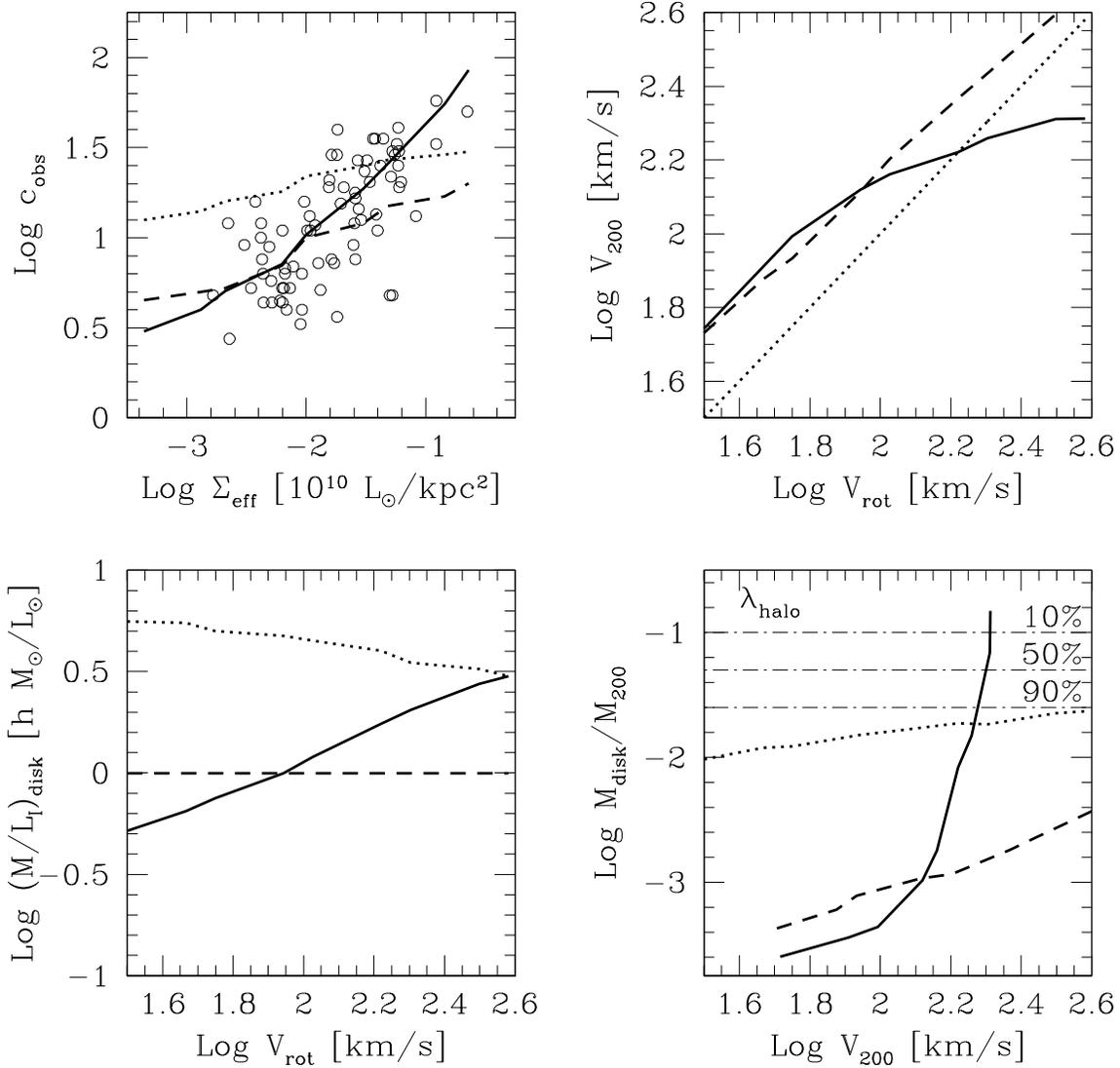}
\figurenum{10} 
\caption{
Top-left: Rotation curve shapes versus effective surface
brightness. The data shown are the same as in Figure 9. We use the
observed exponential scalelengths and luminosities of galaxies in our
sample to compute the shape of the rotation curves expected if these
galaxies are placed in NFW halos with concentrations expected in a
flat, COBE-normalized, $\Omega_0=0.2$ CDM universe ($c \approx 3$, see
bottom line in Figure 7).  Different curves are computed under
different assumptions for the mass-to-light ratio of the disk and for
the circular velocity of the surrounding halos. Dotted lines assume
that $V_{\rm rot}=V_{200}$ and only $(M/L)_{\rm disk}$ is adjusted to match
the Tully-Fisher relation (see Figure 1). Short-dashed lines assume
that $(M/L)_{\rm disk} = h M_{\odot}/L_{\odot}$ in all galaxies; only
$V_{200}$ is varied to match the Tully-Fisher relation. Solid lines
correspond to varying both $(M/L)_{\rm disk}$ {\it and} $V_{200}$ so as to
match the Tully-Fisher relation and the $\Sigma_{\rm eff}$-$c_{\rm obs}$
correlation. The average trends illustrated by the curves have
substantial scatter. The dot-dashed curves labeled $\lambda_{halo}$
in the bottom-right panel indicate the fraction of systems with
rotation parameter higher than shown in the figure. For example,
regardless of mass, only $10\%$ of halos are expected to have
$\lambda_{halo}>0.1$.}
\end{figure}   
\begin{figure} 
\plotone{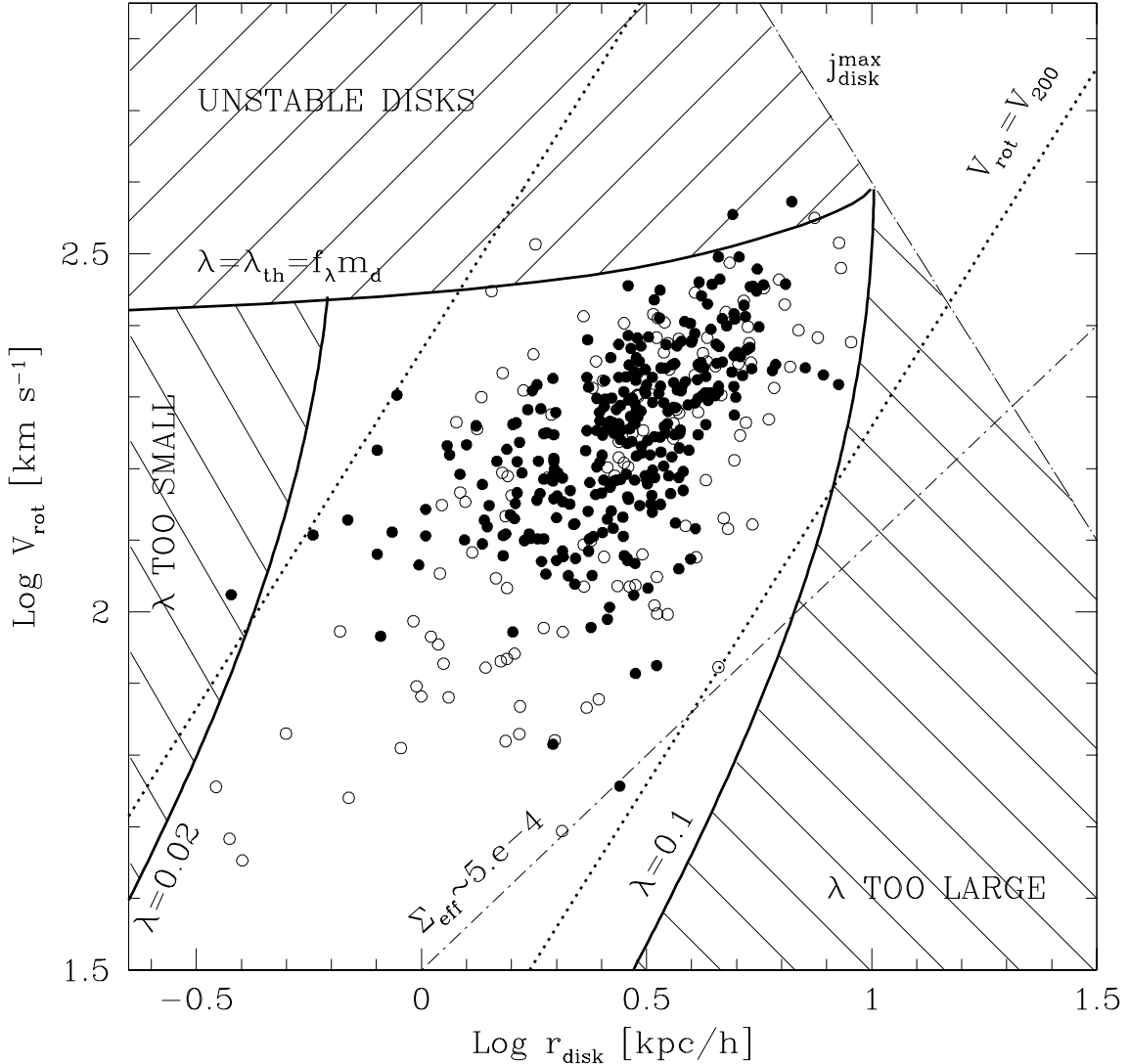}
\figurenum{11} 
\caption{
Distribution of disk-dominated galaxies in the exponential
scalelength-rotation speed plane. Open circles are galaxies in the
sample described in this paper. Filled circles correspond to data
collected by Courteau (1997). Note that most galaxies lie in the wedge
of the plane delimited by the ``maximal-rotators'' (curve labeled
$\lambda=0.1$) and the limit imposed by disk stability arguments
(curve labeled $\lambda=\lambda_{th}$). The dot-dashed line labeled
$j_{\rm disk}^{\rm max}$ illustrates the ``maximum'' angular momentum of halos
with $V_{200}\approx 200$ \kms. Constant effective surface brightness
curves are parallel to the dot-dashed line labeled $\Sigma_{eff}=2
\times 10^{-4}$. ($\Sigma_{eff}$ is quoted in units of $10^{10}
L_{\odot}/$kpc$^2$). Galaxies below this line would be very difficult
to detect because of their extremely low surface brightness. The
meaning of other curves is discussed in the text.}
\end{figure}   

%
\end{document}